\documentclass[12pt]{amsart}

\input xypic
\input xy
\xyoption{all}
\usepackage{graphicx}
\usepackage{amsmath}
\usepackage{amssymb}
\usepackage{amscd}
\usepackage{color}
\usepackage{oldgerm}
\usepackage{amsfonts}
\usepackage{newlfont}
\usepackage{longtable}
\usepackage{multirow}

\usepackage{fontenc}

\DeclareOldFontCommand{\rm}{\normalfont\rmfamily}{\mathrm}

\usepackage{upref}

\theoremstyle{plain}\swapnumbers

\title[The Hurwitz-Hopf Map
and Harmonic Wave Functions for Integer and Half-Integer Angular Momentum]
{The Hurwitz-Hopf Map
and Harmonic Wave Functions for Integer and Half-Integer Angular Momentum}

\author{Sergio A. Hojman${}^{1,2,3;(a)}$, 
Eduardo Nahmad-Achar${}^{4;(b)}$, 
and Adolfo S\'anchez-Valenzuela${}^{5;(c)}$}

\address{(1) Departamento de Ciencias, Facultad de Artes Liberales, Universidad Adolfo Ib\'a\~nez, Santia\-go, Chile.
\hfill\break
(2) Departamento de F\'{\i}sica, Facultad de Ciencias, Universidad de Chile, Santiago, Chile.
\hfill\break
(3) Centro de Recursos Educativos Avanzados, CREA, Santiago, Chile.\hfill\break
(4) Instituto de Ciencias Nucleares, 
Universidad Nacional Aut\'onoma de M\'exico, 
Apdo. Postal 70-543, Mexico City CP 04510, M\'exico.
\hfill\break
(5) Centro de Investigaci\'on	
en Matem\'aticas A.C. Unidad M\'erida
Km 5.5 Carretera Sierra Papacal - Chuburn\'a Puerto. M\'erida, Yucat\'an. CP 97302, M\'exico.}

\email{(A) sergio.hojman@uai.cl; (B) nahmad@nucleares.unam.mx; (C) adolfo@cimat.mx}

\begin{document}

\maketitle

\begin{abstract}
Harmonic wave functions for 
integer and half-integer angular momentum are given
in terms of the Euler angles $(\theta,\phi,\psi)$ that define
a rotation in $SO(3)$, and the Euclidean norm $r$ in ${\mathbb R}^3$,
keeping the usual meaning of the spherical coordinates $(r,\theta,\phi)$.
They form a Hilbert (super)-space decomposed in the form 
$\mathcal H=\mathcal H_0\oplus\mathcal H_1$.
Following a classical work by Schwinger, $2$-dimensional harmonic oscillators
are used to produce raising and lowering operators that change
the total angular momentum eigenvalue of the wave functions in half units.
The nature of the representation space $\mathcal H$ 
is approached from the double covering group homomorphism $SU(2)\to SO(3)$
and the topology involved is taken care of by using 
the Hurwitz-Hopf map $H:{\mathbb R}^4\to{\mathbb R}^3$. 
It is shown how to reconsider $H$ as a 2-to-1 group map, 
$G_0={\mathbb R}^+\times SU(2)\to {\mathbb R}^+\times SO(3)$,
translating it into an assignment $(z_1,z_2)\mapsto (r,\theta,\phi,\psi)$ 
whose domain consists of pairs $(z_1,z_2)$ of complex variables,
under the appropriate identification of ${\mathbb R}^4$ with ${\mathbb C}^2$.
It is shown how the Lie algebra of $G_0$ is coupled with two Heisenberg
Lie algebras of $2$-dimensional (Schwinger's) harmonic oscillators
generated by the operators $\{z_1,z_2,\bar{z}_1,\bar{z}_2\}$ and their adjoints. 
The whole set of operators gets algebraically closed either into a 
$13$-dimensional Lie algebra or into a $(4|8)$-dimensional 
Lie superalgebra.
The wave functions in $\mathcal H$ can be written in terms of
polynomials in the complex coordinates $(z_1,z_2)$
and their complex conjugates $(\bar{z}_1,\bar{z}_2)$
and the representations are explicitly constructed via 
the various highest weight (or lowest weight)
vector representations of $G_0$.
Finally, a new non–relativistic quantum (Schr\"odinger-like) equation
for the hydrogen atom that takes into account the electron spin
is introduced and expressed in terms of $(r,\theta,\phi,\psi)$
and the time $t$.
The equation is succeptible to be solved exactly in terms of the 
harmonic wave functions hereby introduced.
\end{abstract}

\medskip
\noindent
\section*{Introduction and summary of results}

\medskip
\noindent
Following Schwinger’s idea of connecting the ladder operators
for the harmonic oscillator with those for the usual angular momentum,
a realisation is constructed which, besides yielding the latter in terms 
of Euler’s angles, yields new operators which change the total angular 
momentum $j$ by half steps: $\hbar/2$.
This realisation thus gives, in a unified manner, the angular
momentum eigenfunctions with integer and half-integer values.
Schwinger's original work goes back to 1965 (see \cite{6}).
Later on, after the concept of {\it supersymmetry\/}
had been introduced, in 1979 he wrote (see \cite{7}):

\medskip
\noindent
{\lq\lq}
{\it The recent recognition of the existence and possible utility
of transformations between particles of different statistics\/}
[For a mathematically oriented review see~\cite{Corwin75,Freedman76,Deser76}] 
{\it had its origin in the properties of certain two-dimensional dual models,
although the concept was clearly prefigured in the simple unification
of all spins and statistics by means of multispinor sources and fields
\/}~\cite{Schwinger66}
[Offstage voice: {\lq\lq}{\it All right, wise guy! 
Then why didn't you do it first\/?}\,{\rq\rq}]
{\it The question is well put and I cannot answer it\/}
[There is a relevant discussion in J. Hadamard, 
{\lq\lq}The Psychology of Invention in the Mathematical Field,{\rq\rq} 
Princeton Univ. Press, Princeton, 1945],
{\it but it may yet be appropriate, and useful, 
to indicate the elementary kinematical basis for such transformations
and, hopefully, to assist in the redirection of these efforts
toward more modest, physical, goals\/.}{\rq\rq}

\medskip
\noindent
Supersymmetry is a quest for a unified theory of elementary particles 
and their interactions~\cite{Varadarajan04} and, apart from being a 
promise for unified field theories, presents softer divergencies than the 
current models. Its natural language is that of Lie algebras, Lie groups, 
and their representations; the mathematical language describing 
symmetries, which plays a central role in modern mathematical 
physics~\cite{Shun-Jen12}. 
So-called {\it Lie superalgebras}
are a generalisation of Lie algebras, 
which include the notion of a
a {\it super vector space}; i.e., a vector space graded by an additional
parity or degree structure, allowing for the interaction 
(via a {\it Lie super bracket\/}) of elements with equal or different parity: 
the Lie super bracket of two elements with the same parity
leads to an even element, whereas
that of two elements with different parity
leads to an odd element (cf. e.g.~\cite{Corwin75,Jarvis23}).
While no supersymmetric particles have been discovered at the 
present energy scales, experimental observations in trapped 
ion simulators have been reported~\cite{Cai22}, and applications 
in various fields such as optics~\cite{Miri13}, 
quantum optics~\cite{Hirokawa15,Tomka15}, 
quantum chaos~\cite{Efetov99}, quantum simulations~\cite{Gharibyan21}, 
and others, have been devised.

\medskip
\noindent
The purpose of this work is to show how Schwinger's ideas 
produce a Hilbert space (a super vector space, actually)
of harmonic wave functions 
describing both, integer and half-integer angular momentum states
from a unified mathematical point of view 
and having an easy algebraic and physical interpretation.
In fact, the produced wave functions
depend on the Euler angles $(\theta,\phi,\psi)$
that define a rotation in $SO(3)$, and on the Euclidean norm $r$ in ${\mathbb R}^3$,
keeping the usual meaning of the spherical coordinates $(r,\theta,\phi)$.
The harmonic wave functions obtained for integer angular momentum states
can be expressed
---up to normalization--- in the form $r^jY_{jm}(\theta,\phi)$,
with non-negative integral powers of $r$ when the total angular momentum 
$j=n$ takes values in $\mathbb N\cup\{0\}$; in this case, the functions
 $Y_{jm}(\theta,\phi)$ are the usual spherical harmonics
defined on the $2$-sphere embedded in ${\mathbb R}^3$. 
On the other hand, 
the harmonic wave functions obtained for half-integer
angular momentum states are expressed in the form
$r^{j}\,e^{i\psi/2}\,y_{jm}(\theta,\phi)$ with 
$j\in \left(2(\mathbb N\cup\{0\})+1\right)/2$
and the functions $y_{jm}$ (like the $Y_{jm}$'s
for non-negative integer values of $j$)
belonging to the corresponding subspaces $\langle j,m\rangle$ 
spanned by those harmonic functions $f$ satisfying
${\text{\bf L}}f=jf$ and ${\text{\bf L}}_z=mf$ 
with respect to the angular momentum
operators described in Eqs \eqref{(33)} in \S6 
and in {\bf Proposition 7.2} in \S7 below. 
In both cases,  $m\in\{-j, -j+1,\ldots, j-1, j\}$.
{\bf Convention:} Here and in what follows we set $\hbar=1$.

\medskip
\noindent
All the harmonic
wave functions obtained for integer and half-integer angular momentum
states can also be written in terms of two complex variables $(z_1,z_2)$
and their complex conjugates $(\bar{z}_1,\bar{z}_2)$, related to
$(r,\theta,\phi,\psi)$ by,
\begin{equation}\label{(1)}
\aligned
z_1 &= \sqrt{r}\,\operatorname{exp}
\displaystyle{\left(i\,\frac{\psi+\phi}{2}\right)}\,\cos
\displaystyle{\frac{\theta}{2}},
\\
z_2 & = \sqrt{r}\,\operatorname{exp}
\displaystyle{\left(i\,\frac{\psi-\phi}{2}\right)}\,\sin
\displaystyle{\frac{\theta}{2}}.
\endaligned
\end{equation}
The nature of this correspondence is approached from the 
well-known double covering group homomorphism 
$SU(2)\to SO(3)$. The topology involved is taken care of by 
using the {\it Hurwitz-Hopf} map $H:{\mathbb R}^4\to{\mathbb R}^3$
as in Hage-Hassan and Kibler \cite{2}. 
The harmonic wave functions for integer and half-integer angular momentum states
are then obtained from the
corresponding representation theory
of the groups $SU(2)$ and $SO(3)$.
Under appropriate identifications, the domain ${\mathbb R}^4$ of $H$
can be viewed as Hamitlon's quaternions 
$\mathbb H$ which are in turn identified with the domain $\mathbb C^2$
of the complex variables $(z_1,z_2)$.
The relationship between the Schwinger transformation 
and the Hurwitz-Hopf map from $S^3$ to $S^2$ is here given explicitly.

\medskip
\noindent
Our approach consists of changing the point of view of $H$
and consider it as a 2-to-1 group map, 
$G_0={\mathbb R}^+\times SU(2)\to {\mathbb R}^+\times SO(3)$,
which translates into the correspondence
${\mathbb C}^2\ni (z_1,z_2)\mapsto (r,\theta,\phi,\psi)\in {\mathbb R}^+\times SO(3)$,
whenever 
 $\left(\begin{smallmatrix} 
\bar{z}_1/\sqrt{r} & \,\,-\bar{z}_2/\sqrt{r} \\ z_2/\sqrt{r} & \,\,\,\,\, z_1/\sqrt{r}
\end{smallmatrix}\right)\in SU(2)$, and $\vert z_1\vert^2+\vert z_2\vert^2=R^2=r>0$.
In particular, $z_1$ and $z_2$ can be written 
in terms of $(\sqrt{r},\theta/2,\phi/2,\psi/2)$
as in \eqref{(1)} above.
The Lie algebra generators,
$\{\text{\bf L}, \text{\bf L}_x, \text{\bf L}_y, \text{\bf L}_z\}$ 
of $\mathfrak{g}=\operatorname{Lie}(G_0)$
find simple expressions in terms of the monomials $\{z_1,z_2,\bar{z}_1,\bar{z}_2\}$
and the partial derivatives with respect to these variables
(see Eqs \eqref{(33)} in \S6 below).
A direct computation shows that,
\begin{equation}\label{(2)}
{\text{\bf L}_x}^2
+{\text{\bf L}_y}^2
+{\text{\bf L}_z}^2
=
{\text{\bf L}}^2 + \text{\bf L}
+\displaystyle{\frac{r}{4}}\,\Delta,
\end{equation}
where,
\begin{equation}\label{(3)}
\Delta=
\left(
\displaystyle{\frac{\partial^2}{\partial z_1\partial \bar{z}_1}}
+
\displaystyle{\frac{\partial^2}{\partial z_2\partial \bar{z}_2}}
\right)
\end{equation}
is the Laplace operator in the domain ${\mathbb R}^4\simeq{\mathbb C}^2$.
It turns out that the wave functions $f$ produced through our approach
satisfy $\Delta f=0$.

\medskip
\noindent
Moreover, in
terms of the complex variables $(z_1,z_2,\bar{z}_1,\bar{z}_2)$,
the wave functions hereby produced 
are obtained in one of two ways as follows:
either, obtain first the highest weight vector
$\vert\,j\,,+j\,\rangle=(z_1\bar{z}_2)^j$ (if 
$j\in{\mathbb N}\cup\{0\}$),
or $\vert\,j\,,+j\,\rangle=z_1(z_1\bar{z}_2)^{\frac{2j-1}{2}}$ 
(if $j\in \left(2(\mathbb N\cup\{0\})+1\right)/2$),
and then move from it 
by successive application of the ladder operator 
$\text{\bf L}_-=\text{\bf L}_x-i\,\text{\bf L}_y$
which preserves the eigenvalue $j$ but lowers the eigenvalue $m$ 
in one unit at each step, 
until reaching a non-zero scalar multiple of the lowest weight vector $\vert\,j\,,-j\,\rangle$,
characterized by $\text{\bf L}_-\vert\,j\,,-j\,\rangle=0$;
or else, produce first the lowest weight vector 
$\vert\,j\,,-j\,\rangle=(\bar{z}_1z_2)^j$
(if $j\in{\mathbb N}\cup\{0\}$),
or $\vert\,j\,,-j\,\rangle=z_2(\bar{z}_1z_2)^{\frac{2j-1}{2}}$ 
(if $j\in \left(2(\mathbb N\cup\{0\})+1\right)/2$), 
and then move from it 
by successive application of the ladder operator 
$\text{\bf L}_+=\text{\bf L}_x+i\,\text{\bf L}_y$
which preserves the eigenvalue $j$ but raises the eigenvalue $m$ 
in one unit at each step, until reaching
a non-zero scalar multiple of the highest weight vector $\vert\,j\,,+j\,\rangle$,
characterized by $\text{\bf L}_+\vert\,j\,,+j\,\rangle=0$. 

\medskip
\noindent
The whole picture is completed by showing
how the Lie algebra of $G_0$ is coupled with two Heisenberg
Lie algebras of $2$-dimensional harmonic oscillators.
Following Schwinger, the $2$-dimensional harmonic oscillators
are used to produce raising and lowering operators that change
the total angular momentum eigenvalue of the wave functions in half units.
It is explained how the complete set of operators gets closed
either into a $13$-dimensional Lie algebra or into a $(4|8)$-dimensional 
Lie superalgebra.
We mention in passing that mathematically related work in 
a similar direction was done in the mid 70's of the last century  
by M. Kashiwara and M. Vergne following ideas introduced by 
R. Howe (see \cite{3} and \cite{4}). It turns out that
their approach is adapted to a joint action of a given symplectic 
and a given orthogonal group, which is a clear indicative of
supersymmetry (see \cite{Corwin75} and \cite{3}). 

\medskip
\noindent
The Hilbert space where our
wave functions {\it live\/} can be naturally decomposed into a direct sum
$\mathcal H=\mathcal H_0\oplus\mathcal H_1$, where $\mathcal H_0$ contains all the
states having integer angular momentum (or {\it space of bosons\/})
and $\mathcal H_1$ contains all the
states having half-integer angular momentum (or {\it space of fermions\/})
thus providing an explicit realisation of a {\it superspace\/} in which both,
the Lie algebra and the Lie superalgebra are faithfully represented.

\medskip
\noindent
Finally, as an example of our approach, we write in \S13 a
new non\-relativistic quantum (Schr\"odinger-like) equation
in terms of the four dimensional spatial coordinates $(r,\theta,\phi,\psi)$
and the time $t$, using the Hydrogen atom potential energy.
As far as we know, this non-relativistic quantum equation
is new indeed and it takes into account the electron spin.
Besides, it can be solved exactly in terms of the 
new harmonic wave functions hereby introduced.
For the new solutions with half-integer values of $j$, 
we have provided the highest weight wave functions 
$\Psi_{j,+j}\in\langle j,+j\rangle$ together with their energy 
eigenvalues $E_{j,+j}$. The solutions
$\Psi_{j,m}\in\langle j,j-k\rangle$ are obtained, up to a constant, 
via $(\text{\bf L}_-)^k\Psi_{j,+j}$ for non-negative integer values of $k$
satisfying $k\le 2j+1$. 

\medskip
\noindent
\section*{1. Schwinger's Heisenberg-Angular-Momentum Lie Algebra}

\medskip
\noindent
Schwinger's Lie algebra is defined in terms of the
$2$-dimensional quantum harmonic oscilator operators,
$a_x$, $a_y$, and their adjoints, $a_x^\dagger$, $a_y^\dagger$,
satisfying the Heisenberg commutation relations,
\begin{equation}\label{(4)}
[a_x,a_x^\dagger]=1,\qquad [a_y,a_y^\dagger]=1.
\end{equation}
Then define,
\begin{equation}\label{(5)}
\aligned
J_+=a_x^\dagger a_y,&\qquad J_-=a_y^\dagger a_x, 
\\
J_z=\frac{1}{2}(a_x^\dagger a_x - a_y^\dagger a_y),& \qquad
J=\frac{1}{2}(a_x^\dagger a_x + a_y^\dagger a_y).
\endaligned
\end{equation}
It follows that,
\begin{equation}\label{(6)}
\aligned
[J_z,J_{\pm}]=\pm J_{\pm}, &\qquad [J_+,J_-]=2J_z,
\\
[J,J_{\pm}]=0,&\qquad [J,J_z]=0.
\endaligned
\end{equation}
It is not difficult to prove that
\begin{equation}\label{(7)}
\alignedat 3
[J,a_x] & = -\frac{1}{2}\,a_x,\quad&\quad [J,a_x^\dagger] & = \frac{1}{2}\,a_x^\dagger,
\\
[J_z,a_x] &= -\frac{1}{2}\,a_x,\quad&\quad [J_z,a_x^\dagger] & = \frac{1}{2}\,a_x^\dagger,
\\
[J,a_y] &= -\frac{1}{2}\,a_y,\quad&\quad [J,a_y^\dagger] & = \frac{1}{2}\,a_y^\dagger,
\\
[J_z,a_y] & = \frac{1}{2}\,a_y,\quad&\quad [J_z,a_y^\dagger] &= -\frac{1}{2}\,a_y^\dagger,
\endalignedat
\end{equation}
and also that
\begin{equation}\label{(8)}
\alignedat 3
[J_+,a_x] & = -a_y,\quad&\quad
[J_+,a_x^\dagger] & =0,
\\
[J_-,a_x] & =0,\quad&\quad
[J_-,a_x^\dagger] & =a_y^\dagger,
\\
[J_+,a_y] & =0,\quad&\quad
[J_+,a_y^\dagger] & =a_x^\dagger,
\\
[J_-,a_y] & = -a_x,\quad&\quad
[J_-,a_y^\dagger]& =0.
\endalignedat
\end{equation}
We shall give a Lie algebra,
realised in terms of operators adapted to the geometry
and the symmetry involved, 
that fully contains the
subalgebra of the $J$'s and the $a$'s
but, for the sake of completeness in the variables used
to represent the operators, it must include an additional
Heisenberg algebra.

\bigskip

\medskip
\noindent
\section*{2. The Hurwitz-Hopf Map}

\medskip
\noindent
From the point of view of Lie's theory,  the Schwinger 
operators $a_*$ and $a^\dagger_*$
transform representation spaces of the 3-dimensional rotation group $SO(3)$
describing states having a non-negative integer {\it total angular momentum\/} $j$ 
(or {\it spin value\/} $s$; i.e., $j\in\mathbb N\cup\{0\}$),
into representation spaces of its {\it spin group\/} or 
{\it double covering group\/}, $SU(2)$;
that is, into states whose 
{\it total angular momentum\/} is $j\pm 1/2$ (or whose {\it spin value\/} is 
$s\pm 1/2$\/), and viceversa.

\medskip
\noindent
Even though the Lie groups $SU(2)$ and $SO(3)$ are locally
isomorphic, they are topologically different. 
To unveil the spin representations of $SO(3)$ one needs to 
consider the double covering map $SU(2)\to SO(3)$.
One way to take the topology involved into account is as in 
Hage-Hassan and Kibler (see \cite{2}) by considering the Hurwitz-Hopf
map, $H:{\mathbb R}^4\ni u\mapsto H(u)=x\in {\mathbb R}^3$,
given by,
\begin{equation}\label{(9)}
\aligned
x_1&=H_1(u_1,u_2,u_3,u_4)=2(u_1u_3+u_2u_4),\\
x_2&=H_2(u_1,u_2,u_3,u_4)=2(u_2u_3-u_1u_4),\\
x_3&=H_3(u_1,u_2,u_3,u_4)={u_1}^2+{u_2}^2 - {u_3}^2-{u_4}^2,
\endaligned
\end{equation}
which has the property that,
\begin{equation}\label{(10)}
r=\sqrt{{x_1}^2+{x_2}^2+{x_3}^2}
={u_1}^2+{u_2}^2+{u_3}^2+{u_4}^2=R^2.
\end{equation}
Thus, $H$ maps $3$-spheres centered at the origin in ${\mathbb R}^4$,
onto $2$-spheres centered at the origin in ${\mathbb R}^3$. 
The Hurwitz-Hopf map can also be described as a map 
$H:{\mathbb C}^2\to{\mathbb R}^3$, using the complex variables,
\begin{equation}\label{(11)}
z_1=u_1+i\,u_2
\qquad\text{and}\qquad
z_2=u_3+i\,u_4,
\end{equation}
so that,
\begin{equation}\label{(12)}
x_1=z_1\bar{z}_2+\bar{z}_1z_2,\quad
x_2=-i(z_1\bar{z}_2-\bar{z}_1z_2),\quad
x_3=z_1\bar{z}_1-z_2\bar{z}_2,
\end{equation}
that is,
\begin{equation}\label{(13)}
\aligned
x_1&=H_1(z_1,z_2)=2\operatorname{Re}(z_1\bar{z}_2),\\
x_2&=H_2(z_1,z_2)=2\operatorname{Im}(z_1\bar{z}_2),\\
x_3&=H_3(z_1,z_2)=\vert z_1\vert^2 - \vert z_2\vert^2\,.
\endaligned
\end{equation}
One possible symmetry group associated to the 
Hurwitz-Hopf map is the {\it real\/} Lie group,
\begin{equation}\label{(14)}
G = 
\left\{
\begin{pmatrix}
\bar{z}_1 & -\bar{z}_2 \\
z_2 &  \,\,\,z_1
\end{pmatrix}
\,\Big\vert\,
\begin{pmatrix}
z_1\\z_2
\end{pmatrix}
\in\mathbb C^2,
\text{\ \ and\ \ }
R^2=|z_1|^2+|z_2|^2 > 0 
\right\},
\end{equation}
whose connected component to the identity $G_0$ is
isomorphic to ${\mathbb R}^+\times SU(2)$, where ${\mathbb R}^+$
stands for the multiplicative group of positive real numbers,
obtained when $R=\sqrt{r}>0$.
The Lie group $SU(2)$ can be identified 
with the $3$-sphere $|z_1|^2+|z_2|^2=1$ of unit quaternions.
It acts on $3$-spheres in ${\mathbb R}^4\simeq{\mathbb H}$
via left (or right) multiplication in ${\mathbb H}$. 
Thus, the Hurwitz-Hopf
map transforms $3$-spheres in ${\mathbb R}^4$ into $2$-spheres in ${\mathbb R}^3$, 
and  ${\mathbb R}^+$ acts either in ${\mathbb R}^4$ or in ${\mathbb R}^3$ by
changing the spheres radii.

\bigskip

\medskip
\noindent
\section*{3. The Lie group $SU(2)$ and its Lie algebra $\mathfrak{su}(2)$}

\medskip
\noindent
To fix the notation we recall that the
special unitary group $SU(2)$ is defined as,
\begin{equation}\label{(15)}
SU(2)= \left\{ A\in\operatorname{Mat}_{2\times 2}(\mathbb C)\mid
A^{\dagger}A=1\!\!1_2,\text{\ \ and\ \ }\operatorname{det}A=1\right\},
\end{equation}
where $A^{\dagger}=(\bar{A})^t$ stands for the conjugate transpose matrix of $A$
and $1\!\!1_2$ is the $2\times 2$ unit matrix. 
Thus, $A\in \operatorname{Mat}_{2\times 2}(\mathbb C)$
belongs to $SU(2)$ if and only if it is invertible and its inverse $A^{-1}$
es equal to $A^{\dagger}$. That is,
\begin{equation}\label{(16)}
\begin{pmatrix} a&b\\ c&d\end{pmatrix}^{-1}\!\!\!=
\displaystyle{\frac{1}{\det A}}\begin{pmatrix} \,\,d & -b \\ -c & \,\,a\end{pmatrix}
=
\begin{pmatrix} \bar{a}&\bar{c}\\ \bar{b}&\bar{d}\end{pmatrix}
=\begin{pmatrix} a&b\\ c&d\end{pmatrix}^{\dagger}.
\end{equation}
Since $\det A=1$ for any $A\in SU(2)$, it follows that,
\begin{equation}\label{(17)}
d = \bar{a},\qquad b = -\bar{c},\qquad\text{and}\qquad\det A = |a|^2+|c|^2=1.
\end{equation}
Therefore,
\begin{equation}\label{(18)}
SU(2) \simeq \left\{ \begin{pmatrix} \,\,a & -\bar{c} \\ \,\, c & \,\,\,\bar{a}\end{pmatrix}
\in\operatorname{Mat}_{2\times 2}(\mathbb C)\,\Big\vert\ |a|^2+|c|^2=1\right\},
\end{equation}
making clear the fact that $SU(2)$ is actually diffeomorphic to the 
unit $3$-sphere $S^3$ in ${\mathbb R}^4\simeq{\mathbb C}^2$.

\medskip
\noindent
The Lie algebra $\mathfrak{su}(2)$
of $SU(2)$ is the {\it real\/} vector space,
\begin{equation}\label{(19)}
\mathfrak{su}(2)\simeq \left\{ X\in\operatorname{Mat}_{2\times 2}(\mathbb C)\mid
X+X^{\dagger}=0,\text{\ \ and\ \ }\operatorname{Tr}X=0\right\}.
\end{equation}
It is well known that $\mathfrak{su}(2)$ is isomorphic to $\mathbb R^3$ via, 
\begin{equation}\label{(20)}
\mathfrak{su}(2)\ni X \ \Longleftrightarrow\ 
X = i\begin{pmatrix} z & x-iy \\ x+iy & -z\end{pmatrix},\text{\ \ with\ \ }
\begin{pmatrix}
x\\y\\z
\end{pmatrix}\in {\mathbb R}^3.
\end{equation}
A convenient set of linearly independent generators for $\mathfrak{su}(2)$ 
over the real field ${\mathbb R}$ is,
$\{ i\sigma_1, i\sigma_2, i\sigma_3\}$, where, the $\sigma_i$'s are
the Pauli matrices
\begin{equation}\label{(21)}
\sigma_1=\!\begin{pmatrix} 0&1\\1&0\end{pmatrix},
\quad
\sigma_2=\!\begin{pmatrix} 0 & -i \\ i & \,\,\,0\end{pmatrix},
\quad
\sigma_3=\!\begin{pmatrix} 1&\,\,\,0\\0&-1\end{pmatrix},
\end{equation}
and,
\begin{equation}\label{(22)}
[\,i\sigma_1,i\sigma_2]=-2\,i\sigma_3,\quad
[\,i\sigma_2,i\sigma_3]=-2\,i\sigma_1,\quad
[\,i\sigma_3,i\sigma_1]=-2\,i\sigma_2.
\end{equation}
Clearly,
\begin{equation}\label{(23)}
\mathfrak{su}(2)\ni X \ \Rightarrow\ 
\operatorname{det}X = x^2+y^2+z^2.
\end{equation}

\bigskip

\medskip
\noindent
\section*{4. The double covering map $SU(2)\to SO(3)$}

\medskip
\noindent
For each $A\in SU(2)$, the assignment $X\mapsto A\,X\,A^{\dagger}$
defines a transformation $\rho(A):\mathfrak{su}(2)\to \mathfrak{su}(2)$ which preserves
the quadratic form, $\operatorname{det}X = x^2+y^2+z^2$,
since $A^{\dagger}A=1\!\!1_2$ and 
$\operatorname{det}(A\,X\,A^{\dagger})=\operatorname{det}(X\,A^{\dagger}A)
=\operatorname{det}X$. Therefore,
$\rho(A)\in SO(3)$.

\medskip
\noindent
It can be proved (and it is well known) that $\rho$ is a $2$ to $1$ map,
and in fact, a local diffeomorphism. This is usually
expressed by saying that $\rho$ induces the exact sequence of groups,
\begin{equation}\label{(24)}
\{1\!\!1_2\}\hookrightarrow {\mathbb Z}_2\hookrightarrow
SU(2)\twoheadrightarrow SO(3)\twoheadrightarrow \{1\!\!1_3\},
\end{equation}
where, ${\mathbb Z}_2=\{\pm 1\!\!1_2\}$ as a subgroup of $SU(2)$,
and $\rho:SU(2)\twoheadrightarrow SO(3)$ is actually 
{\it a group homomorphism\/,} as it satisfies,
\begin{equation}\label{(25)}
\rho(A_1\, A_2)
 = \rho(A_1)\,\rho(A_2),
 \qquad\text{for any}\ \ A_1,A_2\in SU(2),
\end{equation}
and it is straightforward to rewrite
$\rho(A_1)$ and $\rho(A_2)$ 
as $3\times 3$ matrices with real entries
acting (in that order) on column vectors 
$\left(\begin{smallmatrix}
x\\y\\z
\end{smallmatrix}\right)\in{\mathbb R}^3$,
whenever,
$X=i\left(\begin{smallmatrix} z & x-iy \\ \,\,x+iy & \,-z\end{smallmatrix}\right)\in\mathfrak{su}(2)$.

\medskip
\noindent
{\bf 4.1 Proposition.}
The parametrisation of a given rotation $\rho(A)\in SO(3)$
that takes an initial $x$-$y$-$z$ frame into a final $X$-$Y$-$Z$ frame 
in ${\mathbb R}^3$ in terms of the Euler angles $(\theta,\phi,\psi)$ is obtained from 
the double covering map $SU(2)\ni A\mapsto\rho(A)\in SO(3)$,
by writing $A\in SU(2)$ in the form,
\begin{equation}\label{(26)}
\aligned
A & = 
{\begin{pmatrix} e^{-i\frac{\psi}{2}} & 0 \\ 0 & e^{i\frac{\psi}{2}}\end{pmatrix}}
\begin{pmatrix} \cos\frac{\theta}{2} & -\sin\frac{\theta}{2}
\\ \sin\frac{\theta}{2} & \,\,\,\,\cos\frac{\theta}{2}\end{pmatrix}
{\begin{pmatrix} e^{-i\frac{\phi}{2}} & 0 \\ 0 & e^{i\frac{\phi}{2}}\end{pmatrix}}
\\
& =
{\begin{pmatrix}
e^{-i\,\left(\frac{\psi+\phi}{2}\right)}\cos\frac{\theta}{2} & -e^{-i\,\left(\frac{\psi-\phi}{2}\right)}\sin\frac{\theta}{2} \\
e^{i\,\left(\frac{\psi-\phi}{2}\right)}\sin\frac{\theta}{2} & \,\,\,\,e^{i\,\left(\frac{\psi+\phi}{2}\right)}\cos\frac{\theta}{2} 
\end{pmatrix}},
\endaligned
\end{equation}
so that $\rho(A)=\text{R}_z(\psi)\circ
\text{R}_y(\theta)\circ\text{R}_z(\phi)\in SO(3)$.

\medskip
\noindent
{\bf Proof:} This is well-known and we shall briefly recall
how the result follows from the
following observations:

\medskip
\noindent
{\bf 1.}
Take $A = \left(\begin{smallmatrix} e^{-i\phi} & 0 \\ 0 & e^{i\phi}\end{smallmatrix}\right)\in SU(2)$.
Then,
$$
\rho(A)X = A\,X\,A^{\dagger}= i\begin{pmatrix} z &  e^{-2i\phi}(x-iy) \\ e^{2i\phi}(x+iy) & -z\end{pmatrix},
$$
corresponds to a counterclockwise rotation, $\text{R}_z(2\phi)$,
of the $x$-$y$ plane
(performed around the $z$-axis in $\mathbb R^3$)
by an angle of $2\phi$.

\medskip
\noindent
{\bf 2.}
Take $A = \left(\begin{smallmatrix} \,\,\cos\theta & -\sin\theta \\ \,\,\sin\theta & \,\,\,\cos\theta\end{smallmatrix}\right)\in SU(2)$. 
Then,
$$
\rho(A)X = A\,X\,A^{\dagger}= i\begin{pmatrix} z\cos 2\theta - x\sin 2\theta & z\sin 2\theta + x\cos 2\theta -iy \\
z\sin 2\theta + x\cos 2\theta +iy &  -(z\cos 2\theta - x\sin 2\theta)\end{pmatrix},
$$
corresponds to a counterclockwise rotation, $\text{R}_y(2\theta)$,
of the $z$-$x$ plane
(performed around the $y$-axis in $\mathbb R^3$)
by an angle of $2\theta$.

\medskip
\noindent
Now, in passing from a given $x$-$y$-$z$ (fixed)
frame in ${\mathbb R}^3$ to a (moving) $X$-$Y$-$Z$ frame
(assuming the most general situation in which
the directions defined by $z$ and $Z$ are linearly independent),
we fix first the orientation of the {\it line of nodes\/} $\ell$ defined by the
intersection of the $x$-$y$ plane with the $X$-$Y$ plane, in such a way that
that the positive direction of $\ell$ coincides with the direction of 
$\hat{z}\times \hat{Z}$, where $\hat{z}$ and $\hat{Z}$ are unit vectors
along the corresponding $z$ and $Z$ axes.
Then, the three rotations defined by the Euler angles $(\theta,\phi,\psi)$
to go from the $x$-$y$-$z$ frame to the $X$-$Y$-$Z$ frame, must be performed
as follows: 
\begin{enumerate}
\item
First make a counterclockwise rotation by an angle $\phi$
around the $z$-axis, to take the $x$-axis into the positive direction
of the line of nodes.
\item
Then make a counterclockwise rotation by an angle $\theta$
around the positive direction
of the line of nodes, to make the $z$-axis coincide with the $Z$-axis.
\item
Finally, make a counterclockwise rotation by an angle $\psi$
around the $Z$-axis to make the positive direction of the line of nodes,
coincide with the positive direction defined by the $X$-axis.
\end{enumerate}
\noindent
Observe that
due to the symmetry under rotations in the $x$-$y$ plane,
the positive direction of the line of nodes can be labeled either by
$x^\prime$ or by
$y^\prime$. All that matters is that the {\it auxiliary} $x^\prime$-$y^\prime$-$z$
{\it frame must have\/} the same orientation as the $x$-$y$-$z$ frame does.
We shall adhere to the convention that the positive direction
of the line of nodes is precisely the
auxiliary $y^\prime$-axis, so that the $x^\prime$-axis will have
the direction of $\hat{y}^\prime\times\hat{z}$, where
$\hat{y}^\prime$ and $\hat{z}$ are unit vectors along the positive
direction of the line of nodes and the $z$-axis, respectively.
From these conventions, the result \eqref{(26)} in the statement 
is now evident.
\qed

\bigskip

\medskip
\noindent
\section*{5. The
Hurwitz-Hopf 
Map in Terms of the Four Real
Variables $(r,\theta,\phi,\psi)$}

\medskip
\noindent
Consider the $4$-dimensional {\it real\/} Lie group,
\begin{equation}\label{(27)}
G = 
\left\{
\begin{pmatrix}
\bar{z}_1 & -\bar{z}_2 \\
z_2 & \,\,\,z_1
\end{pmatrix}
\in
\operatorname{Mat}_{2\times 2}(\mathbb C)
\,\Big\vert\,
\ \ 
|z_1|^2+|z_2|^2=R^2 > 0 \ 
\right\}.
\end{equation}
For any $g=\left(\begin{smallmatrix} 
\bar{z}_1 & -\bar{z}_2 \\
z_2 & \,\,\,z_1\end{smallmatrix}\right)\in G$, choose 
a positive real number $R=\sqrt{r}\in{\mathbb R}^+$, so as to have $g$
in {\it the identity component $G_0$ of $G$\/,} and write,
\begin{equation}\label{(28)}
g = \sqrt{r}\begin{pmatrix} 
\bar{z}_1/\sqrt{r} & -\bar{z}_2/\sqrt{r} \\
z_2/\sqrt{r} & \,\,\,z_1/\sqrt{r}\end{pmatrix},
\ \text{with,}\ \ 
\displaystyle{\frac{1}{\sqrt{r}}}
\begin{pmatrix}
\bar{z}_1 & -\bar{z}_2 \\
z_2 & \,\,\,z_1
\end{pmatrix}
\in SU(2).
\end{equation}
Thus, using \eqref{(26)},
$$
\displaystyle{\frac{1}{\sqrt{r}}}
\begin{pmatrix}
\bar{z}_1 & -\bar{z}_2 \\
z_2 & \,\,\,z_1
\end{pmatrix}
= 
{\begin{pmatrix}
e^{-i\,\left(\frac{\psi+\phi}{2}\right)}\cos\frac{\theta}{2} & -e^{-i\,\left(\frac{\psi-\phi}{2}\right)}\sin\frac{\theta}{2} \\
e^{i\,\left(\frac{\psi-\phi}{2}\right)}\sin\frac{\theta}{2} & \,\,\,\,e^{i\,\left(\frac{\psi+\phi}{2}\right)}\cos\frac{\theta}{2} 
\end{pmatrix}}
\in SU(2),
$$
and,
\begin{equation}\label{(29)}
\aligned
z_1 
& = 
\sqrt{r}\,e^{i\,\left(\frac{\psi+\phi}{2}\right)}\cos\frac{\theta}{2},
\\
z_2 
& = 
\sqrt{r}\,e^{i\,\left(\frac{\psi-\phi}{2}\right)}\sin\frac{\theta}{2}.
\endaligned
\end{equation}
In particular, 
using these expressions for $z_1$ and $z_2$,
the cartesian coordinates $x_1$, $x_2$ and $x_3$
---given by the Hurwitz-Hopf map, $x_i=H_i(z_1,z_2)$ ($1\le i\le 3$)--- 
can be written in terms of the spherical coordinates $r$, $\theta$ and $\phi$,
in the usual way,
\begin{equation}\label{(30)}
x_1=r\sin\theta\cos\phi,\qquad
x_2=r\sin\theta\sin\phi,\qquad
x_3=r\cos\theta.
\end{equation}
Also, $H$ yields the following expressions for $r$ and the Euler
angles $(\theta,\phi,\psi)$ in terms of the complex variables
$z_1$ and $z_2$:
\begin{equation}\label{(31)}
\aligned
H^*r & =\vert z_1\vert^2+\vert z_2\vert^2,
\\
H^*\theta & =\cos^{-1}\left(
\displaystyle{\frac{\vert z_1\vert^2-\vert z_2\vert^2}{\vert z_1\vert^2+\vert z_2\vert^2}}
\right),
\\
H^*\phi & =\tan^{-1}\left(
-i\,\displaystyle{\frac{z_1\bar{z}_2-\bar{z}_1z_2}{z_1\bar{z}_2+\bar{z}_1z_2}}
\right),
\\
H^*\psi & =
\tan^{-1}\left(-i\,\displaystyle{\frac{z_1z_2 - \bar{z}_1\bar{z}_2}{z_1z_2 + \bar{z}_1\bar{z}_2}}
\right),
\endaligned
\end{equation}
where, $H^*$ stands for {\it the pullback of\/} $H$, so that
$H^*f$ means $f\circ H$ for any smooth function $f$. Thus, for example,
$H^*r=r\left(H(z_1,z_2)\right)=\vert z_1\vert^2+\vert z_2\vert^2$, etc.
In particular, $H$ itself gets described as a map changing two
specific local coordinate charts; 
say, $(z_1,z_2)\mapsto (r,\theta,\phi,\psi)$,
thus allowing a reinterpretation of the Hurwitz-Hopf map
as a $2$-to-$1$ differentiable group map,
\begin{equation}\label{(32)}
{\aligned
H:{\Bbb R}^+\times SU(2) & \rightarrow {\Bbb R}^+\times SO(3)
\\
(z_1,z_2) &\mapsto (r,\theta,\phi,\psi),\qquad |z_1|^2+|z_2|^2=r>0.
\endaligned}
\end{equation}

\medskip
\noindent
\section*{6. The Lie algebra $\mathfrak{g}$ of the 
Symmetry Group $G$ associated to the Hurwitz-Hopf Map
}

\medskip
\noindent
Since $G_0\simeq {\mathbb R}^+\times SU(2)$, it is immediate to see that
the Lie algebra $\mathfrak{g}$ of $G$ is isomorphic to the {\it real\/} 
$4$-dimensional space,
\begin{equation}\label{(33)}
\mathfrak{g}=\operatorname{Span}_{\mathbb R}\{\,1\!\!1_2\,\}\oplus\mathfrak{su}(2).
\end{equation}
Its complexification $\mathfrak{g}_{\mathbb C}$ is isomporphic to the
Lie algebra $\mathfrak{gl}_2(\mathbb C)$ of complex $2\times 2$ matrices.
Geometrically, we may
realize the Lie algebras $\mathfrak{g}$ and $\mathfrak{g}_{\mathbb C}$
in terms of real and complex $4$-dimensional
vector spaces of left $G$-invariant vector fields, respectively.
A convenient set of linearly independent 
generators for either $\mathfrak{g}$ or  $\mathfrak{g}_{\mathbb C}$, is
$\{\text{\bf L},\text{\bf L}_z,\text{\bf L}_+,\text{\bf L}_-\}$, where,
\begin{equation}\label{(34)}
\aligned
\text{\bf L} & = \displaystyle{\frac{1}{2}}\,\left(\,
z_1
\displaystyle{\frac{\partial}{\partial z_1}}
+z_2
\displaystyle{\frac{\partial}{\partial z_2}}
+\bar{z}_1
\displaystyle{\frac{\partial}{\partial \bar{z}_1}}
+\bar{z}_2
\displaystyle{\frac{\partial}{\partial \bar{z}_2}}
\,\right),
\\
\text{\bf L}_z & = \displaystyle{\frac{1}{2}}\,\left(\,
z_1
\displaystyle{\frac{\partial}{\partial z_1}}
-z_2
\displaystyle{\frac{\partial}{\partial z_2}}
-\bar{z}_1
\displaystyle{\frac{\partial}{\partial \bar{z}_1}}
+\bar{z}_2
\displaystyle{\frac{\partial}{\partial \bar{z}_2}}
\,\right),
\\
\text{\bf L}_+ & = \left(\,
z_1
\displaystyle{\frac{\partial}{\partial z_2}}
-\bar{z}_2
\displaystyle{\frac{\partial}{\partial \bar{z}_1}}
\,\right),
\qquad\text{and,}
\\
\text{\bf L}_- & = -\left(\,
\bar{z}_1
\displaystyle{\frac{\partial}{\partial \bar{z}_2}}
-z_2
\displaystyle{\frac{\partial}{\partial z_1}}
\,\right).
\endaligned
\end{equation}
It is a straightforward matter to verify that $\text{\bf L}$ commutes
with all the operators and that,
\begin{equation}\label{(35)}
[\text{\bf L}_z,\text{\bf L}_\pm]=\pm\text{\bf L}_\pm,
\qquad\text{and}\qquad
[\text{\bf L}_+,\text{\bf L}_-]=2\text{\bf L}_z,
\end{equation}
thus proving that $\operatorname{Span}_{\mathbb C}\{
\text{\bf L},\text{\bf L}_z,\text{\bf L}_+,\text{\bf L}_-\}$
is isomorphic to the Lie algebra
$\mathfrak{gl}_2(\mathbb C)$. Clearly, the operators \eqref{(34)}
may act on the space of complex 
polynomials in the variables $\{z_1,z_2,\bar{z}_1,\bar{z}_2\}$
producing polynomials in the same variables.

\medskip
\noindent
\section*{7. Computations with the Angular Momentum Complex Operators}

\medskip
\noindent
Take the Lie algebra generators 
$\{\text{\bf L},\text{\bf L}_z,\text{\bf L}_+,\text{\bf L}_-\}$  
as given before, and define the operators 
$\text{\bf L}_x$ and $\text{\bf L}_y$ through,
\begin{equation}\label{(36)}
\text{\bf L}_+ =
\text{\bf L}_x+i\,\text{\bf L}_y,
\qquad\text{and}\qquad
\text{\bf L}_- =
\text{\bf L}_x-i\,\text{\bf L}_y.
\end{equation}
Therefore,
\begin{equation}\label{(37)}
\text{\bf L}_x =\displaystyle{\frac{1}{2}} \left(\,\text{\bf L}_+ + \text{\bf L}_-\,\right),
\qquad\text{and}\qquad
\text{\bf L}_y =\displaystyle{\frac{1}{2i}} \left(\,\text{\bf L}_+ - \text{\bf L}_-\,\right).
\end{equation}
It turns out that,
\begin{equation}\label{(38)}
[\text{\bf L}_x,\text{\bf L}_y]=i\,\text{\bf L}_z,\qquad
[\text{\bf L}_z,\text{\bf L}_x]=i\,\text{\bf L}_y,\qquad
[\text{\bf L}_y,\text{\bf L}_z]=i\,\text{\bf L}_x.
\end{equation}
Clearly,
\begin{equation}\label{(39)}
\text{\bf L}_+\text{\bf L}_- = {\text{\bf L}_x}^2+ {\text{\bf L}_y}^2-i\,[\text{\bf L}_x,\text{\bf L}_y]
={\text{\bf L}_x}^2+ {\text{\bf L}_y}^2 +\text{\bf L}_z\,,
\end{equation}
whence,
\begin{equation}\label{(40)}
{\text{\bf L}_x}^2+ {\text{\bf L}_y}^2+{\text{\bf L}_z}^2
=\text{\bf L}_+\text{\bf L}_- +{\text{\bf L}_z}^2 - \text{\bf L}_z\,.
\end{equation}

\medskip
\noindent
{\bf 7.1 Proposition.} A direct computation shows that,
$$
{\text{\bf L}_x}^2
+{\text{\bf L}_y}^2
+{\text{\bf L}_z}^2
=
{\text{\bf L}}^2 + \text{\bf L}
+\left(
\vert z_1\vert^2+\vert z_2\vert^2
\right)
\left(
\displaystyle{\frac{\partial^2}{\partial z_1\partial \bar{z}_1}}
+
\displaystyle{\frac{\partial^2}{\partial z_2\partial \bar{z}_2}}
\right),
$$
where, 
$$
\left(
\displaystyle{\frac{\partial^2}{\partial z_1\partial \bar{z}_1}}
+
\displaystyle{\frac{\partial^2}{\partial z_2\partial \bar{z}_2}}
\right)
=
\Delta
=
\displaystyle{\frac{1}{4}}\,
\left(
\displaystyle{\frac{\partial^2}{{\partial u_1}^2}}+
\displaystyle{\frac{\partial^2}{{\partial u_2}^2}}+
\displaystyle{\frac{\partial^2}{{\partial u_3}^2}}+
\displaystyle{\frac{\partial^2}{{\partial u_4}^2}}
\right),
$$
is the Laplace operator in ${\mathbb R}^4$.

\medskip
\noindent
{\bf Proof:} This is a straightforward computation.
One only has to make use of the 
explicit expressions for $\text{\bf L}_\pm$ and $\text{\bf L}_z$
given in \eqref{(34)} and apply the operator
$\text{\bf L}_+\text{\bf L}_- +{\text{\bf L}_z}^2 - \text{\bf L}_z$
appearing on the right hand side of \eqref{(40)} 
to an arbitrary differentiable function $f$
depending on the local coordinates $(z_1,z_2,\bar{z}_1,\bar{z}_2)$.
The calculation shows that
$(\text{\bf L}_+\text{\bf L}_- +{\text{\bf L}_z}^2 - \text{\bf L}_z)f$ is
equal to $(\text{\bf L}^2+\text{\bf L}+(|z_1|^2+|z_2|^2)\Delta)f$
where $\Delta$ is the Laplace operator in the statement. 
\qed

\medskip
\noindent
{\bf 7.2 Proposition.}
Using the expressions
$H^*r$, $H^*\theta$, $H^*\phi$ and $H^*\psi$ given in \eqref{(31)},
together with the covering map \eqref{(32)} stating that
$$
{
\aligned
z_1 
& = 
\sqrt{r}\,e^{i\,\left(\frac{\psi+\phi}{2}\right)}\cos\frac{\theta}{2},
\\
z_2 
& = 
\sqrt{r}\,e^{i\,\left(\frac{\psi-\phi}{2}\right)}\sin\frac{\theta}{2},
\endaligned}
\ \ 
\text{whenever}
\ \ 
\begin{pmatrix}
\bar{z}_1 & -\bar{z}_2 \\
z_2 & \,\,\, z_1
\end{pmatrix}
\in
G_0=
{\mathbb R}^+\times SU(2),
$$
then
$$
\displaystyle{\frac{\partial}{\partial z_1}}
=
\sqrt{r}\,e^{-i\,\left(\frac{\psi+\phi}{2}\right)}\,
\left(
\cos{\frac{\theta}{2}}\,
\displaystyle{\frac{\partial}{\partial r}}
-\displaystyle{\frac{1}{r}}
\,\sin\frac{\theta}{2}\,
\displaystyle{\frac{\partial}{\partial \theta}}
-\displaystyle{\frac{i}{2r\,\cos{\frac{\theta}{2}}}}
{\left(
\displaystyle{\frac{\partial}{\partial\phi}}
+
\displaystyle{\frac{\partial}{\partial\psi}}
\right)}
\right),
$$
$$
\displaystyle{\frac{\partial}{\partial z_2}}
=
\sqrt{r}\,e^{-i\,\left(\frac{\psi-\phi}{2}\right)}\,
\left(
\sin{\frac{\theta}{2}}\,
\displaystyle{\frac{\partial}{\partial r}}
+\displaystyle{\frac{1}{r}}
\,\cos\frac{\theta}{2}\,
\displaystyle{\frac{\partial}{\partial \theta}}
+\displaystyle{\frac{i}{2r\,\sin{\frac{\theta}{2}}}}
{\left(
\displaystyle{\frac{\partial}{\partial\phi}}
-
\displaystyle{\frac{\partial}{\partial\psi}}
\right)}
\right),
$$
$$
\displaystyle{\frac{\partial}{\partial \bar{z}_1}}
=
\sqrt{r}\,e^{i\,\left(\frac{\psi+\phi}{2}\right)}\,
\left(
\cos{\frac{\theta}{2}}\,
\displaystyle{\frac{\partial}{\partial r}}
-\displaystyle{\frac{1}{r}}
\,\sin\frac{\theta}{2}\,
\displaystyle{\frac{\partial}{\partial \theta}}
+\displaystyle{\frac{i}{2r\,\cos{\frac{\theta}{2}}}}
{\left(
\displaystyle{\frac{\partial}{\partial\phi}}
+
\displaystyle{\frac{\partial}{\partial\psi}}
\right)}
\right),
$$
$$
\displaystyle{\frac{\partial}{\partial \bar{z}_2}}
=
\sqrt{r}\,e^{i\,\left(\frac{\psi-\phi}{2}\right)}\,
\left(
\sin{\frac{\theta}{2}}\,
\displaystyle{\frac{\partial}{\partial r}}
+\displaystyle{\frac{1}{r}}
\,\cos\frac{\theta}{2}\,
\displaystyle{\frac{\partial}{\partial \theta}}
-\displaystyle{\frac{i}{2r\,\sin{\frac{\theta}{2}}}}
{\left(
\displaystyle{\frac{\partial}{\partial\phi}}
-
\displaystyle{\frac{\partial}{\partial\psi}}
\right)}
\right).
$$
Moreover,
$$
\text{\bf L} = \displaystyle{\frac{1}{2}}\,\left(\,
z_1
\displaystyle{\frac{\partial}{\partial z_1}}
+z_2
\displaystyle{\frac{\partial}{\partial z_2}}
+\bar{z}_1
\displaystyle{\frac{\partial}{\partial \bar{z}_1}}
+\bar{z}_2
\displaystyle{\frac{\partial}{\partial \bar{z}_2}}
\,\right) = r\,\displaystyle{\frac{\partial}{\partial r}},
$$
$$
\text{\bf L}_z = \displaystyle{\frac{1}{2}}\,\left(\,
z_1
\displaystyle{\frac{\partial}{\partial z_1}}
-z_2
\displaystyle{\frac{\partial}{\partial z_2}}
-\bar{z}_1
\displaystyle{\frac{\partial}{\partial \bar{z}_1}}
+\bar{z}_2
\displaystyle{\frac{\partial}{\partial \bar{z}_2}}
\,\right)=
-i\,\displaystyle{\frac{\partial}{\partial\phi}},
$$
$$
\text{\bf L}_+ 
=
\left(\,
z_1
\displaystyle{\frac{\partial}{\partial z_2}}
-\bar{z}_2
\displaystyle{\frac{\partial}{\partial \bar{z}_1}}
\,\right)
=
e^{i\phi}\,
\left(
\displaystyle{\frac{\partial}{\partial \theta}}
+i\,\displaystyle{\frac{\cos\theta}{\sin\theta}}
{\left(
\displaystyle{\frac{\partial}{\partial\phi}}
-
\displaystyle{\frac{1}{\cos\theta}}\,
\displaystyle{\frac{\partial}{\partial\psi}}
\right)}
\right),
$$
$$
\text{\bf L}_- 
=
-\left(\,
\bar{z}_1
\displaystyle{\frac{\partial}{\partial \bar{z}_2}}
-z_2
\displaystyle{\frac{\partial}{\partial z_1}}
\,\right)
=
-e^{-i\phi}\,
\left(
\displaystyle{\frac{\partial}{\partial \theta}}
-i\,\displaystyle{\frac{\cos\theta}{\sin\theta}}
{\left(
\displaystyle{\frac{\partial}{\partial\phi}}
-
\displaystyle{\frac{1}{\cos\theta}}\,
\displaystyle{\frac{\partial}{\partial\psi}}
\right)}
\right).
$$

\medskip
\noindent
{\bf Proof:} 
The operators $\partial/\partial{z_1}$, $\partial/\partial{z_2}$,
$\partial/\partial{\bar{z}_1}$ and $\partial/\partial{\bar{z}_2}$
can be written in terms of $\partial/\partial r$, $\partial/\partial\theta$, 
$\partial/\partial\phi$ and $\partial/\partial\psi$, by using the chain rule:
$$
\displaystyle{\frac{\partial}{\partial z_1}}=
\displaystyle{\frac{\partial H^*r}{\partial z_1}}\,
\displaystyle{\frac{\partial}{\partial r}} +
\displaystyle{\frac{\partial H^*\theta}{\partial z_1}}\,
\displaystyle{\frac{\partial}{\partial \theta}} +
\displaystyle{\frac{\partial H^*\phi}{\partial z_1}}\,
\displaystyle{\frac{\partial}{\partial \phi}} +
\displaystyle{\frac{\partial H^*\psi}{\partial z_1}}\,
\displaystyle{\frac{\partial}{\partial \psi}},
$$
$$
\displaystyle{\frac{\partial}{\partial z_2}}=
\displaystyle{\frac{\partial H^*r}{\partial z_2}}\,
\displaystyle{\frac{\partial}{\partial r}} +
\displaystyle{\frac{\partial H^*\theta}{\partial z_2}}\,
\displaystyle{\frac{\partial}{\partial \theta}} +
\displaystyle{\frac{\partial H^*\phi}{\partial z_2}}\,
\displaystyle{\frac{\partial}{\partial \phi}} +
\displaystyle{\frac{\partial H^*\psi}{\partial z_2}}\,
\displaystyle{\frac{\partial}{\partial \psi}},
$$
$$
\displaystyle{\frac{\partial}{\partial \bar{z}_1}}=
\displaystyle{\frac{\partial H^*r}{\partial \bar{z}_1}}\,
\displaystyle{\frac{\partial}{\partial r}} +
\displaystyle{\frac{\partial H^*\theta}{\partial \bar{z}_1}}\,
\displaystyle{\frac{\partial}{\partial \theta}} +
\displaystyle{\frac{\partial H^*\phi}{\partial \bar{z}_1}}\,
\displaystyle{\frac{\partial}{\partial \phi}} +
\displaystyle{\frac{\partial H^*\psi}{\partial \bar{z}_1}}\,
\displaystyle{\frac{\partial}{\partial \psi}},
$$
$$
\displaystyle{\frac{\partial}{\partial \bar{z}_2}}=
\displaystyle{\frac{\partial H^*r}{\partial \bar{z}_2}}\,
\displaystyle{\frac{\partial}{\partial r}} +
\displaystyle{\frac{\partial H^*\theta}{\partial \bar{z}_2}}\,
\displaystyle{\frac{\partial}{\partial \theta}} +
\displaystyle{\frac{\partial H^*\phi}{\partial \bar{z}_2}}\,
\displaystyle{\frac{\partial}{\partial \phi}} +
\displaystyle{\frac{\partial H^*\psi}{\partial \bar{z}_2}}\,
\displaystyle{\frac{\partial}{\partial \psi}}.
$$
The rest are just straightforward
calculations of the partial derivatives of $H^*r$, $H^*\theta$, $H^*\phi$
and $H^*\psi$ with respect to $z_1$, $z_2$, $\bar{z}_1$ and $\bar{z}_2$ 
using the expressions in the statement.
\qed

\medskip
\noindent
{\bf 7.3 Corollary.}
Since
$$
\text{\bf L}_x=\displaystyle{\frac{1}{2}}\left(\text{\bf L}_+ + \text{\bf L}_-\right),
\qquad\text{and}\qquad
\text{\bf L}_y=\displaystyle{\frac{1}{2i}}\left(\text{\bf L}_+ - \text{\bf L}_-\right),
$$
it follows from {\bf Prop. 7.2} that,
$$
\text{\bf L}_x=
i\,\left(
\sin\phi\,\displaystyle{\frac{\partial}{\partial \theta}}
+\cos\phi\,\displaystyle{\frac{\cos\theta}{\sin\theta}}
{\left(
\displaystyle{\frac{\partial}{\partial\phi}}
-
\displaystyle{\frac{1}{\cos\theta}}\,
\displaystyle{\frac{\partial}{\partial\psi}}
\right)}
\right),
$$
and
$$
\text{\bf L}_y=
-i\,\left(
\cos\phi\,\displaystyle{\frac{\partial}{\partial \theta}}
-\sin\phi\,\displaystyle{\frac{\cos\theta}{\sin\theta}}
{\left(
\displaystyle{\frac{\partial}{\partial\phi}}
-
\displaystyle{\frac{1}{\cos\theta}}\,
\displaystyle{\frac{\partial}{\partial\psi}}
\right)}
\right).\qed
$$


\medskip
\noindent
{\bf 7.4 Remark.}
It is worthwhile to observe the form of the slightly 
simplified expressions just obtained 
for the angular momentum operators $\text{\bf L}_\pm$ and $\text{\bf L}_z$,
as well as the expression that results
for ${\text{\bf L}_x}^2+{\text{\bf L}_y}^2+{\text{\bf L}_z}^2$; namely,
\begin{equation}\label{(41)}
\aligned
\text{\bf L}_+ & =\operatorname{e}^{i\phi}
\left(
\displaystyle{\frac{\partial}{\partial \theta}} + i\,\cot \theta\,
\displaystyle{\frac{\partial}{\partial \phi}} 
-i\,\displaystyle{\frac{1}{\sin\theta}}
\displaystyle{\frac{\partial}{\partial\psi}}
\right),
\\
\text{\bf L}_- & =\operatorname{e}^{-i\phi}
\left(
\displaystyle{\frac{\partial}{\partial \theta}} - i\,\cot \theta\,
\displaystyle{\frac{\partial}{\partial \phi}} 
+i\,\displaystyle{\frac{1}{\sin\theta}}
\displaystyle{\frac{\partial}{\partial\psi}}
\right),
\\
\text{\bf L}_z & = -i\,\displaystyle{\frac{\partial}{\partial \phi}},
\endaligned
\end{equation}
\begin{equation}\label{(42)}
\aligned
{\text{\bf L}_x}^2+{\text{\bf L}_y}^2+{\text{\bf L}_z}^2 & = 
-\left(
\displaystyle{\frac{\partial^2}{\partial\theta^2}}+
\cot\theta\,\displaystyle{\frac{\partial}{\partial\theta}}+
\csc\theta\,\displaystyle{\frac{\partial^2}{\partial\phi^2}}
\right)
\\
&\quad
-\left(
\csc^2\theta\,\displaystyle{\frac{\partial^2}{\partial\psi^2}}
-2\,\csc\theta\,\displaystyle{\frac{\partial}{\partial\phi}}
\displaystyle{\frac{\partial}{\partial\psi}}
\right).
\endaligned
\end{equation}
which coincide with the usual ones, except for the terms 
depending on $\partial/\partial\psi$. 
Observe that such terms vanish when acting on the usual
spherical wave functions depending on $(r,\theta,\phi)$.
 Nevertheless, they are crucial when acting
on spherical harmonics half-integer spin wave functions
(see \S11, \S12 and \S13 below).
At any rate, it is important to remark that these new angular momentum
operators satisfy the usual angular momentum commutation relations,
as they are simply a particular realisation of the Lie algebra
$\mathfrak{su}_2\simeq\mathfrak{so}_3$.

\medskip
\noindent
\section*{8. The Complete Set of Schwinger's Operators}

\medskip
\noindent
Following Schwinger's works \cite{6} and \cite{7},
introduce the following set of {\it eight operators\/}:
\begin{equation}\label{(43)}
\alignedat 4
\text{\bf j}_1^{++} & = z_1,
\qquad
\text{\bf j}_1^{+-} & = \bar{z}_1,
\qquad
\text{\bf j}_1^{-+} & = \displaystyle{\frac{\partial}{\partial z_2}},
\qquad
\text{\bf j}_1^{--} & = \displaystyle{\frac{\partial}{\partial z_1}},
\\
\text{\bf j}_2^{++} & = \bar{z}_2,
\qquad
\text{\bf j}_2^{+-} & = z_2,
\qquad
\text{\bf j}_2^{-+} & = \displaystyle{\frac{\partial}{\partial \bar{z}_1}},
\qquad
\text{\bf j}_2^{--} & = \displaystyle{\frac{\partial}{\partial \bar{z}_2}}.
\endalignedat
\end{equation}
We shall refer to them collectively as {\lq\lq}{\it the\/} 
$\text{\bf j}_*^{**}$-{\it operators\/}{\rq\rq},
or simply as {\lq\lq}{\it the\/} $\text{\bf j}$'s{\rq\rq}.
Just like the operators from the set 
$\operatorname{Span}_{\Bbb C}\{\text{\bf L},\text{\bf L}_z,\text{\bf L}_+,
\text{\bf L}_-\}\simeq\mathfrak{gl}_2({\mathbb C})$
---to which we shall refer as the {\lq\lq}{\it the\/} $\text{\bf L}_*$-operators{\rq\rq}, 
or simply as {\it the\/} $\text{\bf L}$'s---
the $\text{\bf j}$'s also act on the space of 
polynomials in the variables $(z_1,z_2,\bar{z}_1,\bar{z}_2)$, to produce
polynomials in the same variables.

 \medskip
 \noindent
{\bf  8.1 Proposition.}
The commutators of the $\text{\bf L}_*$-operators with the 
$\text{\bf j}_*^{**}$-operators, produce $\text{\bf j}_*^{**}$-operators. 
In other words, under commutators between $\text{\bf L}$'s and $\text{\bf j}$'s,
the $\text{\bf j}$'s form a representation
space for the Lie algebra $\mathfrak{gl}_2(\mathbb C)$ spanned by
the $\text{\bf L}$'s.
Moreover, the subspace generated by the
$\text{\bf j}$'s gets decomposed into four copies
of the $2$-dimensional irreducible representation 
of $\mathfrak{gl}_2({\mathbb C})$ according to:

\medskip
\noindent
{\bf 1.}
On $\operatorname{Span}\{\text{\bf j}_1^{++},\text{\bf j}_2^{+-}\}$
$$
\text{\bf L} = \displaystyle{\frac{1}{2}}\begin{pmatrix}
1 & 0 \\ 0 & 1
\end{pmatrix},
\quad
\text{\bf L}_z = \displaystyle{\frac{1}{2}}\begin{pmatrix}
1 & 0 \\ 0 & \!\!\!-1
\end{pmatrix},
\quad
\text{\bf L}_+ = \begin{pmatrix}
0 & 1 \\ 0 & 0
\end{pmatrix},
\quad
\text{\bf L}_- = \begin{pmatrix}
0 & 0 \\ 1 & 0
\end{pmatrix}.
$$
\medskip
\noindent
{\bf 2.}
On $
\operatorname{Span}\{\text{\bf j}_1^{-+},\text{\bf j}_1^{--}\}$,
$$
\text{\bf L} = -\displaystyle{\frac{1}{2}}\begin{pmatrix}
1 & 0 \\ 0 & 1
\end{pmatrix},
\quad
\text{\bf L}_z = \displaystyle{\frac{1}{2}}\begin{pmatrix}
1 & 0 \\ 0 & \!\!\!-1
\end{pmatrix},
\quad
\text{\bf L}_+ = -\begin{pmatrix}
0 & 1 \\ 0 & 0
\end{pmatrix},
\quad
\text{\bf L}_- = -\begin{pmatrix}
0 & 0 \\ 1 & 0
\end{pmatrix}.
$$
\medskip
\noindent
{\bf 3.}
On $\operatorname{Span}\{\text{\bf j}_2^{++},\text{\bf j}_1^{+-}\}$,
$$
\text{\bf L} = \displaystyle{\frac{1}{2}}\begin{pmatrix}
1 & 0 \\ 0 & 1
\end{pmatrix},
\quad
\text{\bf L}_z = \displaystyle{\frac{1}{2}}\begin{pmatrix}
1 & 0 \\ 0 & \!\!\!-1
\end{pmatrix},
\quad
\text{\bf L}_+ = -\begin{pmatrix}
0 & 1 \\ 0 & 0
\end{pmatrix},
\quad
\text{\bf L}_- = -\begin{pmatrix}
0 & 0 \\ 1 & 0
\end{pmatrix}.
$$
\medskip
\noindent
{\bf 4.}
On 
$\operatorname{Span}\{\text{\bf j}_2^{-+},\text{\bf j}_2^{--}\}$,
$$
\text{\bf L} = -\displaystyle{\frac{1}{2}}\begin{pmatrix}
1 & 0 \\ 0 & 1
\end{pmatrix},
\quad
\text{\bf L}_z = \displaystyle{\frac{1}{2}}\begin{pmatrix}
1 & 0 \\ 0 & \!\!\!-1
\end{pmatrix},
\quad
\text{\bf L}_+ = \begin{pmatrix}
0 & 1 \\ 0 & 0
\end{pmatrix},
\quad
\text{\bf L}_- = \begin{pmatrix}
0 & 0 \\ 1 & 0
\end{pmatrix}.
$$

\medskip 
\noindent
{\bf Proof:} This is a straightforward calculation using
\eqref{(34)} and \eqref{(43)}. The results can be just listed as follows:
$$
 [\text{\bf L},\text{\bf j}_1^{++} ] =  \displaystyle{\frac{1}{2}}\,\text{\bf j}_1^{++},
\quad
 [\text{\bf L},\text{\bf j}_1^{+-} ] = \displaystyle{\frac{1}{2}}\,\text{\bf j}_1^{+-},
\quad
 [\text{\bf L},\text{\bf j}_1^{-+} ] = -\displaystyle{\frac{1}{2}}\,\text{\bf j}_1^{-+},
\quad
 [\text{\bf L},\text{\bf j}_1^{--} ] = -\displaystyle{\frac{1}{2}}\,\text{\bf j}_1^{--},
$$

$$
 [\text{\bf L},\text{\bf j}_2^{++} ] = \displaystyle{\frac{1}{2}}\,\text{\bf j}_2^{++},
\quad
 [\text{\bf L},\text{\bf j}_2^{+-} ] = \displaystyle{\frac{1}{2}}\,\text{\bf j}_2^{+-},
\quad
 [\text{\bf L},\text{\bf j}_2^{-+} ] = -\displaystyle{\frac{1}{2}}\,\text{\bf j}_2^{-+},
\quad
 [\text{\bf L},\text{\bf j}_2^{--} ] = -\displaystyle{\frac{1}{2}}\,\text{\bf j}_2^{--},
$$

$$
 [\text{\bf L}_z,\text{\bf j}_1^{++} ] =  \displaystyle{\frac{1}{2}}\,\text{\bf j}_1^{++},
\quad
 [\text{\bf L}_z,\text{\bf j}_1^{+-} ] = - \displaystyle{\frac{1}{2}}\,\text{\bf j}_1^{+-},
\quad
 [\text{\bf L}_z,\text{\bf j}_1^{-+} ] = \displaystyle{\frac{1}{2}}\,\text{\bf j}_1^{-+},
\quad
 [\text{\bf L}_z,\text{\bf j}_1^{--} ] = -\displaystyle{\frac{1}{2}}\,\text{\bf j}_1^{--},
$$

$$
 [\text{\bf L}_z,\text{\bf j}_2^{++} ] =  \displaystyle{\frac{1}{2}}\,\text{\bf j}_2^{++},
\quad
 [\text{\bf L}_z,\text{\bf j}_2^{+-} ] =  -\displaystyle{\frac{1}{2}}\,\text{\bf j}_2^{+-},
\quad
 [\text{\bf L}_z,\text{\bf j}_2^{-+} ] = \displaystyle{\frac{1}{2}}\,\text{\bf j}_2^{-+},
\quad
 [\text{\bf L}_z,\text{\bf j}_2^{--} ] = -\displaystyle{\frac{1}{2}}\,\text{\bf j}_2^{--},
$$

$$
 [\text{\bf L}_+,\text{\bf j}_1^{++} ] = 0,
\quad
 [\text{\bf L}_+,\text{\bf j}_1^{+-} ] =- \text{\bf j}_2^{++},
\quad
 [\text{\bf L}_+,\text{\bf j}_1^{-+} ] =  0,
\quad
 [\text{\bf L}_+,\text{\bf j}_1^{--} ] =  -\text{\bf j}_1^{-+},
$$

$$
 [\text{\bf L}_+,\text{\bf j}_2^{++} ] = 0,
\quad
 [\text{\bf L}_+,\text{\bf j}_2^{+-} ] = \text{\bf j}_1^{++},
\quad
 [\text{\bf L}_+,\text{\bf j}_2^{-+} ] = 0,
\quad
 [\text{\bf L}_+,\text{\bf j}_2^{--} ] = \text{\bf j}_2^{-+},
$$

$$
 [\text{\bf L}_-,\text{\bf j}_1^{++} ] = \text{\bf j}_2^{+-},
\quad
 [\text{\bf L}_-,\text{\bf j}_1^{+-} ] = 0,
\quad
 [\text{\bf L}_-,\text{\bf j}_1^{-+} ] = -\text{\bf j}_1^{--},
\quad
 [\text{\bf L}_-,\text{\bf j}_1^{--} ] = 0,
$$

$$
 [\text{\bf L}_-,\text{\bf j}_2^{++} ] =  -\text{\bf j}_1^{+-},
\quad
 [\text{\bf L}_-,\text{\bf j}_2^{+-} ] = 0,
\quad
 [\text{\bf L}_-,\text{\bf j}_2^{-+} ] = \text{\bf j}_2^{--},
\quad
 [\text{\bf L}_-,\text{\bf j}_2^{--} ] = 0.
 \qed
$$

\medskip
\noindent
We write
\begin{equation}\label{(44)}
\aligned
W_{1}
& = W_{1}^{+1/2}\oplus W_{1}^{-1/2} =
\operatorname{Span}\{\text{\bf j}_1^{++},\text{\bf j}_2^{+-}\}\oplus
\operatorname{Span}\{\text{\bf j}_1^{-+},\text{\bf j}_1^{--}\}\,,
\\
W_{2}
& = W_{2}^{+1/2}\oplus W_{2}^{-1/2} = 
\operatorname{Span}\{\text{\bf j}_2^{++},\text{\bf j}_1^{+-}\}\oplus
\operatorname{Span}\{\text{\bf j}_2^{-+},\text{\bf j}_2^{--}\}\,.
\endaligned
\end{equation}
According to {\bf Prop. 8.1}, the four $2$-dimensional subspaces
involved in these direct sums
are $2$-dimensional irreducible representations for $\mathfrak{gl}_2(\mathbb C)$ on which 
the operator $\text{\bf L}$ acts as the scalar $j=\pm 1/2$,
as indicated by the superscript in the notation $W_i^{\pm 1/2}$
($1\le i\le 2$).
On each of these $2$-dimensional subspaces, the operator $\text{\bf L}_z$
acts diagonally in the given ordered bases as $\operatorname{diag}(+1/2,-1/2)$.
Both, $W_{1}$ and $W_{2}$ are symplectic vector spaces and the given decompositions
$W_{i}^{+1/2}\oplus W_{i}^{-1/2}$ 
correspond to the totally isotropic $2$-dimensional subspaces 
defined by their corresponding symplectic forms
with which one may define $5$-dimensional Heisenberg Lie algebras
as in \cite{6} and \cite{7} (see \S 9 below).

\medskip
\noindent
{\bf 8.2 Definition.}
Let $\rho:\mathfrak{gl}_2(\mathbb C)\to \mathfrak{gl}(W_1\oplus W_2)$
be the representation defined by the Lie brackets computed in
{\bf Prop. 8.1}. That, is, for any $x\in \mathfrak{gl}_2(\mathbb C)$ and any 
$w\in W_1\oplus W_2$, $\rho(x)(w)=[x,w]$.

\medskip
\noindent
{\bf 8.3 Corollary.} One
may deduce the following possible correspondences
with the original Schwinger operators \eqref{(4)} and \eqref{(5)}
given in \S1. Either,
$$
\aligned
a_x^\dagger & \leftrightarrow \text{\bf j}_1^{++},
\qquad
a_y^\dagger \leftrightarrow \text{\bf j}_2^{+-},
\qquad
a_x \leftrightarrow \text{\bf j}_1^{--},
\qquad
a_y  \leftrightarrow \text{\bf j}_1^{-+},
\quad\text{or,}
\\
a_x^\dagger & \leftrightarrow \text{\bf j}_2^{++},
\qquad
a_y^\dagger \leftrightarrow \text{\bf j}_1^{+-},
\qquad
a_x \leftrightarrow \text{\bf j}_2^{--},
\qquad
a_y \leftrightarrow \text{\bf j}_2^{-+}.\qquad\qed
\endaligned
$$

\medskip
\noindent
\section*{9. The Lie algebra vs the Lie superalgebra approach}

\medskip
\noindent
Assuming that the operators $\text{\bf L}_*$ and $\text{\bf j}_*^{**}$
act on an appropriate Hilbert space of functions, $\mathcal H$, one may consider
the {\bf Lie algebra},
\begin{equation}\label{(45)}
\mathfrak{g} = \langle \operatorname{Id}\rangle\oplus\mathfrak{gl}_2\oplus W_{1}\oplus W_{2},
\end{equation}
where $\langle \operatorname{Id}\rangle$ stands for the one-dimensional 
subspace generated by the identity operator $\operatorname{Id}:\mathcal H\to\mathcal H$.
In this case, the subspaces,
\begin{equation}\label{(46)}
\mathfrak{h}_{1}=
W_{1}\oplus  \langle \operatorname{Id}\rangle,
\qquad\text{and}\qquad
\mathfrak{h}_{2}=
W_{2}\oplus  \langle \operatorname{Id}\rangle,
\end{equation}
close under the Lie bracket $[\,\cdot\,,\,\cdot\,]$ 
which is the ordinary commutator of the input operators,
and each $\mathfrak{h}_{*}$ is a $5$-dimensional Heisenberg algebra.
Moreover, $[\mathfrak{h}_{1},\mathfrak{h}_{2}]=0$,
and $\mathfrak{h}_{1}\cap\mathfrak{h}_{2}=\langle \operatorname{Id}\rangle$.
Observe that one requires $W_{1}$ and $W_{2}$
to be symplectic spaces 
in order to close the algebraic structure via Lie brackets (i.e., commutators)
between $\text{\bf j}$'s, giving thus rise to the two Heisenberg Lie algebras \eqref{(46)}.

\medskip
\noindent
On the other hand, one may consider the {\bf Lie superalgebra},
\begin{equation}\label{(47)}
\mathfrak{g}/ \langle \operatorname{Id}\rangle \simeq
\mathfrak{g}^\prime = 
 \mathfrak{gl}_2\oplus W_{1}\oplus W_{2},
\end{equation}
where the algebraic structure 
for pairs of elements in $W_{1}\oplus W_{2}$
gets closed through a {\bf symmetric bilinear $\mathfrak{gl}_2(\mathbb C)$-valued}
map $\Gamma:(W_1\oplus W_2)\times (W_1\oplus W_2)\to \mathfrak{gl}_2(\mathbb C)$,
satisfying the equivariance condition,
\begin{equation}\label{(48)}
[x,\Gamma(v,w)] = \Gamma\left(\rho(x)(v),w\right)+\Gamma\left(v,\rho(x)(w)\right),
\end{equation}
for the representation $\rho$ defined in {\bf 8.2} and
for any $x\in \mathfrak{gl}_2(\mathbb C)$, and any pair $v$ and $w$ in $W_1\oplus W_2$.
The underlying $\mathbb Z_2$-graded vector space of this Lie superalgebra is,
\begin{equation}\label{(49)}
\mathfrak{g}^\prime =\mathfrak{g}^\prime_0\oplus \mathfrak{g}^\prime_1,
\qquad
\text{where,}
\qquad
\begin{cases}
&\!\!\!\!\mathfrak{g}^\prime_0 = \mathfrak{gl}_2, \\
&\!\!\!\!\mathfrak{g}^\prime_1 = W_{1}\oplus W_{2},
\end{cases}
\end{equation}
and the Lie superbracket is computed as,
\begin{equation}\label{(50)}
[x+v,y+w]=[x,y]+\rho(x)(w)-\rho(y)(v)+\Gamma(v,w),
\end{equation}
for any pair $x$ and $y$ in $\mathfrak{gl}_2(\mathbb C)$
and any pair $v$ and $w$ in  
$\mathfrak{g}^\prime_1 =W_{1}\oplus W_{2}$ (see \cite{Corwin75} and \cite{5}).
Moreover, since $\mathfrak{g}^\prime_1$
may be decomposed in terms of
various copies of the fundamental $2$-dimensional irreducible representation
of $\mathfrak{g}^\prime_0 = \mathfrak{gl}_2$,
we may apply the results from \cite{5} to conclude the following: 

\medskip
\noindent
{\bf 9.1 Proposition.} For the Lie superalgebra just described through
\eqref{(47)} to \eqref{(50)}, any symmetric bilinear
$\rho$-equivariant $\mathfrak{gl}_2(\mathbb C)$-valued
map $\Gamma:(W_1\oplus W_2)\times (W_1\oplus W_2)
\to \mathfrak{gl}_2(\mathbb C)$, must be identically zero.\qed

\medskip
\noindent
{\bf 9.2 Remark.} For the Lie superalgebra \eqref{(49)} with Lie superbracket \eqref{(50)},
the parity of all the the generators of $W_1\oplus W_2$ (i.e., the $\text{\bf j}$-operators
\eqref{(43)}) is $1$; that is, they are {\it odd operators\/.} Thus, in order to build up
a Lie supergroup whose Lie superalgebra is \eqref{(49)}, one would have to consider first
the $GL_2$-homogeneous vector bundle defined over $GL_2$ itself, whose typical fiber
is the odd supspace $W_1\oplus W_2$, and then look at the supermanifold based on 
$GL_2$ whose defining supermanifold sheaf is the sheaf of sections of the
exterior algebra ($GL_2$-homogeneous) bundle $\wedge(W_1\oplus W_2)$.
In particular, the Lie superbracket $\Gamma(u,v)$ of any two odd generators is symmetric,
and when viewed as sections of the corresponding supermanifold sheaf, they anti-commute
(see ~\cite{Peniche-SV} or ~\cite{Rothstein}).

\medskip
\noindent
\section*{10. The Representation Space and the Wave Functions}

\medskip
\noindent
For either the Lie algebra $\mathfrak{g}$ or the Lie superalgebra $\mathfrak{g}^\prime$,
we shall take the representation space to be the $L^2$-completion
with respect to the hermitian product given by,
\begin{equation}\label{(51)}
\langle \,\psi_1\,,\,\psi_2\,\rangle =\int_{{\mathbb R}^4}
\psi_1(z_1,z_2)\,\overline{\psi_2(z_1,z_2)}\,
e^{-\frac{1}{2}(\vert z_1\vert^2+\vert z_2\vert^2)}\,
dz_1\,dz_2\,d\bar{z}_1\,d\bar{z}_2,
\end{equation}
in the vector space generated by the harmonic polynomials
that result ---up to normalization constants---
from  $(\text{\bf L}_+)^k(\bar{z}_1z_2)^n$, 
together with those that result from
$(\text{\bf L}_+)^kz_2(\bar{z}_1z_2)^n$
($0\le k\le n$ and $n\in\{0\}\cup {\mathbb N}$).
It will be clear from our work in \S11 and \S12 below,
that these expressions are linear combinations
of monomials of the form
${\,\bar{z}_2}^k{z_1}^{m-k}$ and
${\,\bar{z}_1}^k{z_2}^{m-k}$ ($0\le k\le m$, $m=n$ or $m=n+1$, 
and $n\in \{0\}\cup{\mathbb N}$), and these are harmonic
as they satisfy, 
$\Delta f=0$, where $\Delta$ is the Laplace operator
given in {\bf Prop. 7.1}.

%
%

\medskip
\noindent
\section*{11. Examples}

\medskip
\noindent
Define $\vert 0,0\rangle =1$
and leave the normalization questions aside, for the moment.
Define the $j=1/2$ kets by,
\begin{equation}\label{(52)}
\aligned
\vert 1/2,+1/2\rangle&=\text{\bf j}_1^{++}\vert 0,0\rangle = z_1\vert 0,0\rangle=z_1, \\
\vert 1/2,-1/2\rangle&=\text{\bf j}_2^{+-}\vert 0,0\rangle = z_2\vert 0,0\rangle=z_2,
\endaligned
\end{equation}
to get the $j=1/2$ {\it harmonic functions\/,}
\begin{equation}\label{(53)}
\aligned
f_{1/2,+1/2}(r,\theta,\phi,\psi)
& = 
\sqrt{r}\,\,e^{i\frac{\psi}{2}}\,e^{i\frac{\phi}{2}}\,\cos\frac{\theta}{2},
\\
f_{1/2,-1/2}(r,\theta,\phi,\psi)
& = 
\sqrt{r}\,\,e^{i\frac{\psi}{2}}\,e^{-i\frac{\phi}{2}}\,\sin\frac{\theta}{2}.
\endaligned
\end{equation}
In terms of the complex coordinates $(z_1,z_2,\bar{z}_1,\bar{z}_2)$, we have,
\begin{equation}\label{(54)}
\aligned
f_{1/2,+1/2}(z_1,z_2,\bar{z}_1,\bar{z}_2)
& = z_1,
\\
f_{1/2,-1/2}(z_1,z_2,\bar{z}_1,\bar{z}_2)
& = z_2.
\endaligned
\end{equation}
Using the $\text{\bf L}$'s as given in \eqref{(34)}, it follows from \eqref{(54)} that,
\begin{equation}\label{(55)}
\text{\bf L}f_{1/2,+1/2} =\displaystyle{\frac{1}{2}}\,f_{1/2,+1/2},
\quad\text{and}\quad
\text{\bf L}f_{1/2,-1/2} =\displaystyle{\frac{1}{2}}\,f_{1/2,-1/2},
\end{equation}
whereas,
\begin{equation}\label{(56)}
\text{\bf L}_zf_{1/2,+1/2} =\displaystyle{\frac{1}{2}}\,f_{1/2,+1/2},
\quad\text{and}\quad
\text{\bf L}_zf_{1/2,-1/2} =-\displaystyle{\frac{1}{2}}\,f_{1/2,-1/2}.
\end{equation}
On the other hand,
\begin{equation}\label{(57)}
\text{\bf L}_- z_1
= z_2,\quad\text{and}\quad
\text{\bf L}_+ z_2= z_1.
\end{equation}
That is,
\begin{equation}\label{(58)}
\aligned
\text{\bf L}_-
f_{1/2,+1/2} & = f_{1/2,-1/2};\quad \text{i.e.,}\quad 
\text{\bf L}_-\vert 1/2,+1/2\rangle = \vert 1/2,-1/2\rangle,\\
\text{\bf L}_+
f_{1/2,-1/2} & = f_{1/2,+1/2};\quad \text{i.e.,}\quad 
\text{\bf L}_+\vert 1/2,-1/2\rangle = \vert 1/2,+1/2\rangle.
\endaligned
\end{equation}
Observe that the pair 
$\left(\begin{smallmatrix} z_1\\z_2\end{smallmatrix}\right)
=\left(\begin{smallmatrix} f_{1/2,+1/2}\\f_{1/2,-1/2}\end{smallmatrix}\right)$
corresponds to the $2$-dimensional {\it spin representation\/.}
If required, this spinor may be acted upon by the {\it gauge transformation\/,}
$\left(\begin{smallmatrix} z_1\\z_2\end{smallmatrix}\right)
\mapsto \,e^{-i\frac{\psi}{2}}\left(\begin{smallmatrix} z_1\\z_2\end{smallmatrix}\right)$
to eliminate its dependence on $\psi$.

\medskip
\noindent
{\bf 11.1. Notation:} 
Let $\langle j,m\rangle$ stand for the subspace
generated by the wave functions $|\,j,m\,\rangle$ ---also denoted by 
$f_{j,m}$--- satisfying,
\begin{equation}\label{(59)}
\text{\bf L}\,|\,j,m\,\rangle = j\,|\,j,m\,\rangle,
\quad\text{and}\quad
\text{\bf L}_z\,|\,j,m\,\rangle = m\,|\,j,m\,\rangle.
\end{equation}
In what follows, we shall use the well known relations
\begin{equation}\label{(60)}
\aligned
\text{\bf L}_+\,|\,j,m\,\rangle 
& =
\sqrt{(j-m)(j+m+1)}\,|\,j,m+1\,\rangle,
\\
\text{\bf L}_-\,|\,j,m\,\rangle 
& =
\sqrt{(j+m)(j-m+1)}\,|\,j,m-1\,\rangle.
\endaligned
\end{equation}

\medskip
\noindent
{\bf 11.2. Proposition.} 
\begin{equation}
\text{\bf j}_1^{+-}\langle 1/2,-1/2\rangle\subset\langle 1,-1\rangle,
\quad\text{and,}\quad
\text{\bf j}_2^{++}\langle 1/2,+1/2\rangle\subset\langle 1,+1\rangle.
\end{equation}

\medskip
\noindent
{\bf Proof:} Since $\langle 1/2,-1/2\rangle=\langle z_2\rangle$, it follows that,
$\text{\bf j}_1^{+-}z_2=\bar{z}_1z_2$. Therefore,
using \eqref{(34)} one obtains that,
$$
\text{\bf L}\,(\bar{z}_1z_2)=\bar{z}_1z_2,
\qquad\text{and}\qquad
\text{\bf L}_z\,(\bar{z}_1z_2)=-\bar{z}_1z_2.
$$
Similarly, since $\langle 1/2,+1/2\rangle=\langle z_1\rangle$, it follows that,
$\text{\bf j}_2^{++}z_1=z_1\bar{z}_2$, and,
$$
\text{\bf L}\,(z_1\bar{z}_2)=z_1\bar{z}_2,
\qquad\text{and}\qquad
\text{\bf L}_z\,(z_1\bar{z}_2)=z_1\bar{z}_2.
\qed
$$ 

\medskip
\noindent
{\bf 11.3. Proposition.}
$$
\aligned
\vert 1,-1\rangle =
\text{\bf j}_1^{+-}z_2 =
\text{\bf j}_1^{+-}\vert 1/2,-1/2\rangle & =
\text{\bf j}_1^{+-}\left(\,\text{\bf j}_2^{+-}\vert 0,0\rangle\right) =
\bar{z}_1z_2,
\\
\vert 1,+1\rangle =
\text{\bf j}_2^{++}z_1 =
\text{\bf j}_2^{++}\vert 1/2,+1/2\rangle & =
\text{\bf j}_2^{++}\left(\,\text{\bf j}_1^{++}\vert 0,0\rangle\right)  =
z_1\bar{z}_2.
\endaligned
$$
{\bf Proof:} Straightforward computations using \eqref{(34)}
show that,
$$
\text{\bf L}(\bar{z}_1z_2)
= 
\bar{z}_1z_2
\ \ \Rightarrow\ \ j=+1,
\quad\text{and}\quad
\text{\bf L}(z_1\bar{z}_2)
= 
z_1\bar{z}_2
\ \ \Rightarrow\ \ j=+1,
$$
$$
\text{\bf L}_z (\bar{z}_1z_2) = - \bar{z}_1z_2
\ \ \Rightarrow\ \ m=-1,
\quad\text{and}\quad
\text{\bf L}_z (z_1\bar{z}_2) = z_1\bar{z}_2
\ \ \ \ \Rightarrow\ \ m=+1.
$$
Straightforward computations also show that,
$$
\text{\bf L}_+ (\bar{z}_1z_2)=
\vert z_1\vert^2 -\vert z_2\vert^2,
\quad
\text{\bf L}_+\left(\vert z_1\vert^2 -\vert z_2\vert^2\right)
=
-2z_1\bar{z}_2,
\quad
\text{\bf L}_+ (-2z_1\bar{z}_2)
=0,
$$
$$
\text{\bf L}_- (z_1\bar{z}_2)
= \vert z_2\vert^2 -\vert z_1\vert^2,
\quad
\text{\bf L}_-\left(\vert z_2\vert^2 -\vert z_1\vert^2\right)
=-2\bar{z}_1z_2,
\quad
\text{\bf L}_-(-2\bar{z}_1z_2)
=0. \quad\qed
$$
In particular, 
$$
\aligned
\vert z_1\vert^2-\vert z_2\vert^2=r\cos\theta & = \text{\bf L}_+\,|\,1,-1\,\rangle = \sqrt{2}\,|\,1,0\,\rangle
\\
&
\Longrightarrow\quad
|\,1,0\,\rangle = \displaystyle{\frac{r}{\sqrt{2}}}\cos\theta,
\\
-2z_1\bar{z}_2=-re^{-i\phi}\sin\theta 
& =(\text{\bf L}_+)^2|\,1,-1\,\rangle 
=
\sqrt{2}\,\text{\bf L}_+\,|\,1,0\,\rangle=2\,|\,1,-1\,\rangle
\\
&\Longrightarrow\quad
|\,1,-1\,\rangle = -\displaystyle{\frac{r}{2}}e^{-i\phi}\sin\theta.
\endaligned
$$
Therefore, we may write,
\begin{equation}\label{(62)}
\begin{pmatrix}
f_{1,+1}(r,\theta,\phi)\\
f_{1,0}(r,\theta,\phi)\\
f_{1,-1}(r,\theta,\phi)
\end{pmatrix}
=
\begin{pmatrix}
r/2\,e^{i\phi}\,\sin\theta\\
r/\sqrt{2}\cos\theta\\
-r/2\,e^{-i\phi}\,\sin\theta
\end{pmatrix}.
\end{equation}
Observe that up to an overall scalar multiple of $r$, the triple
$\left(\begin{smallmatrix} f_{1,+1}\\f_{1,0}\\f_{1,-1}\end{smallmatrix}\right)$
of the spin-one representation
yields the ordinary spherical harmonics
$\left(\begin{smallmatrix} 
Y_{1,+1}(\theta,\phi)\\Y_{1,0}(\theta,\phi)\\Y_{1,-1}(\theta,\phi)\end{smallmatrix}\right)$
as expected.

\medskip
\noindent
{\bf 11.4. Summary.} 
One may summarize the results obtained in passing from $j=1/2$
to $j=1$ by means of the following {\lq}starting{\rq} diagram: 
\begin{equation}\label{(63)}
\diagram
{\langle 1,-1\rangle}\rto^{\text{\bf L}_+}&{\langle 1,0\rangle}&
{\langle 1,+1\rangle}\lto_{\text{\bf L}_-}\\
\uto^{\text{\bf j}_1^{+-}}
{\langle 1/2,-1/2\rangle}\urto_{\alpha}
&& 
\ulto^{\beta}
{\langle 1/2,+1/2\rangle}\uto_{\text{\bf j}_2^{++}}\\
&\ulto^{\text{\bf j}_2^{+-}}{\langle 0,0\rangle}\urto_{\text{\bf j}_1^{++}}|>\tip & 
\enddiagram
\end{equation}
where, the maps $\alpha$ and $\beta$ can be determined
using the commutation relations between the $\text{\bf L}$'s 
and the $\text{\bf j}^{**}_{*}$'s:
\begin{equation}\label{(64)}
\aligned
\alpha=
\text{\bf L}_+\circ\text{\bf j}_1^{+-}&=
-\text{\bf j}_2^{++} +\text{\bf j}_1^{+-}\circ\text{\bf L}_+,
\\
\beta=
\text{\bf L}_-\circ\text{\bf j}_2^{++}&=
-\text{\bf j}_1^{+-} +\text{\bf j}_2^{++}\circ\text{\bf L}_-.
\endaligned
\end{equation}

\bigskip

\medskip
\noindent
\section*{
12. Generalisation}

\medskip
\noindent
{\bf 12.1. Proposition.} 
For the $j=n$ or the $j=n+1/2$ 
($n\in \mathbb N\cup\{0\}$)
spin representation, the corresponding lowest weight and highest weight
vectors, respectively characterized by $\text{\bf L}_-\vert\,j\,,-j\,\rangle=0$
and $\text{\bf L}_+\vert\,j\,,+j\,\rangle=0$, are given by
$$
\aligned
\vert \,n,-n\rangle & = 
\left(\text{\bf j}_1^{+-}\text{\bf j}_2^{+-}\right)^n\,\vert 0,0\rangle = 
(\bar{z}_1{z_2})^n
\\
\vert \,n,+n\rangle & = 
\left(\text{\bf j}_2^{++}\text{\bf j}_1^{++}\right)^n\,\vert 0,0\rangle =
 ({z_1}\bar{z}_2)^n
 \\
\vert \,n+1/2,-(n+1/2)\rangle & = \text{\bf j}_2^{+-}
\left(\text{\bf j}_1^{+-}\text{\bf j}_2^{+-}\right)^n\,\vert 0,0\rangle = 
z_2(\bar{z}_1{z_2})^n
\\
\vert \,n+1/2,+(n+1/2)\rangle & = \text{\bf j}_1^{++}
\left(\text{\bf j}_2^{++}\text{\bf j}_1^{++}\right)^n\,\vert 0,0\rangle =
z_1({z_1}\bar{z}_2)^n
\endaligned
$$

\medskip
\noindent
{\bf Proof:} Straightforward computations
using the $\text{\bf L}$'s given in \eqref{(34)}, show indeed that,
$$
\aligned
\text{\bf L}(\bar{z}_1z_2)^n
& = 
n\,(\bar{z}_1z_2)^n,
\\
\text{\bf L}({\bar{z}_1}^n{z_2}^{n+1}) 
& = 
\left(n+\displaystyle{\frac{1}{2}}\right){\bar{z}_1}^n{z_2}^{n+1},
\\
\text{\bf L}_z(\bar{z}_1z_2)^n 
& = 
-n\,(\bar{z}_1z_2)^n,
\\
\text{\bf L}_z({\bar{z}_1}^n{z_2}^{n+1}) 
&=
-\left(n+\displaystyle{\frac{1}{2}}\right){\bar{z}_1}^n{z_2}^{n+1},
\endaligned
$$
and
$$
\aligned
\text{\bf L}(z_1\bar{z}_2)^n 
& = 
n\,(z_1\bar{z}_2)^n,
\\
\text{\bf L}({z_1}^{n+1}{\bar{z}_2}^n) 
& = 
\left(n+\displaystyle{\frac{1}{2}}\right){z_1}^{n+1}{\bar{z}_2}^n,
\\
\text{\bf L}_z(z_1\bar{z}_2)^n 
& = 
+n\,(z_1\bar{z}_2)^n,
\\
\text{\bf L}_z({z_1}^{n+1}{\bar{z}_2}^n) 
& = 
+\left(n+\displaystyle{\frac{1}{2}}\right){z_1}^{n+1}{\bar{z}_2}^n.
\endaligned
$$
To complete the proof one needs to show that,
$$
\text{\bf L}_-(\bar{z}_1z_2)^n=0,\quad\text{and}\quad
\text{\bf L}_-({\bar{z}_1}^n{z_2}^{n+1})=0,
$$
and that,
$$
\text{\bf L}_+(z_1\bar{z}_2)^n=0,\quad\text{and}\quad
\text{\bf L}_+({z_1}^{n+1}{\bar{z}_2}^n)= 0.
$$
The general cases of these results, need induction together with
the statements of {\bf Lemma 12.2} and {\bf Cor. 12.3} below.

\medskip
\noindent
{\bf 12.2. Lemma.}
$$
\aligned
\text{\bf L}_+\left(\text{\bf j}_1^{+-}\text{\bf j}_2^{+-}\right)
&=\left(\text{\bf j}_1^{+-}\text{\bf j}_2^{+-}\right)\text{\bf L}_+
+\left(\text{\bf j}_1^{+-}\text{\bf j}_1^{++}-\text{\bf j}_2^{++}\text{\bf j}_2^{+-}\right)
\\
&=\left(\text{\bf j}_1^{+-}\text{\bf j}_2^{+-}\right)\text{\bf L}_+
+\left(\,\vert z_1\vert^2 - \vert z_2\vert^2\right)\operatorname{Id}
\\
\text{\bf L}_-\left(\text{\bf j}_2^{++}\text{\bf j}_1^{++}\right)
&=\left(\text{\bf j}_2^{++}\text{\bf j}_1^{++}\right)\text{\bf L}_-
+\left(\text{\bf j}_2^{++}\text{\bf j}_2^{+-}-\text{\bf j}_1^{+-}\text{\bf j}_1^{++}\right)
\\
&=\left(\text{\bf j}_2^{++}\text{\bf j}_1^{++}\right)\text{\bf L}_-
+\left(\,\vert z_2\vert^2-\vert z_1\vert^2\right)\operatorname{Id}
\endaligned
$$

 \medskip
\noindent
{\bf Proof:} This is a straightforward calculation using the
expressions for the $\text{\bf L}$'s in \eqref{(34)} and those
for the $\text{\bf j}$'s in \eqref{(43)}, together with the
commutation relations of {\bf Prop. 8.1}.\qed

\medskip
\noindent
{\bf 12.3. Corolary.} Since $[\text{\bf L}_+,\text{\bf j}_2^{+-}]=\text{\bf j}_1^{++}$
and $[\text{\bf L}_-,\text{\bf j}_1^{++}]=\text{\bf j}_2^{+-}$, it follows that,
$$ 
\text{\bf L}_+\text{\bf j}_2^{+-}
 =
\text{\bf j}_2^{+-}\text{\bf L}_+ +\text{\bf j}_1^{++},
$$
implying that,
$$
\text{\bf L}_+\,\text{\bf j}_2^{+-}\left(\text{\bf j}_1^{+-}\text{\bf j}_2^{+-}\right)
=\text{\bf j}_2^{+-}\left(\text{\bf j}_1^{+-}\text{\bf j}_2^{+-}\right)\text{\bf L}_+
+\left(\vert z_2\vert^2\,z_2\right)\operatorname{Id}.
$$

\medskip
\noindent
{\bf Proof:} Indeed,
$$
\aligned
\text{\bf L}_+\,\text{\bf j}_2^{+-}\left(\text{\bf j}_1^{+-}\text{\bf j}_2^{+-}\right)
& =
\text{\bf j}_2^{+-}\text{\bf L}_+\left(\text{\bf j}_1^{+-}\text{\bf j}_2^{+-}\right)
+ \text{\bf j}_1^{++}\left(\text{\bf j}_1^{+-}\text{\bf j}_2^{+-}\right)
\\
& = 
\left(\text{\bf j}_2^{+-}\text{\bf j}_1^{+-}\text{\bf j}_2^{+-}\right)\text{\bf L}_+
+\left(\,\vert z_2\vert^2-\vert z_1\vert^2\right)\text{\bf j}_2^{+-}
+\vert z_1\vert^2\text{\bf j}_2^{+-}
\\
&=\left(\text{\bf j}_2^{+-}\text{\bf j}_1^{+-}\text{\bf j}_2^{+-}\right)\text{\bf L}_+
+\vert z_2\vert^2\,\text{\bf j}_2^{+-}
\\
&=\text{\bf j}_2^{+-}\left(\text{\bf j}_1^{+-}\text{\bf j}_2^{+-}\right)\text{\bf L}_+
+\left(\vert z_2\vert^2\,z_2\right)\operatorname{Id},
\endaligned
$$
as claimed.\qed

%
%

\medskip
\noindent
{\bf 12.4. Corollary.} For each non-negative integer $n$, the
components $f$ of the $2n +1$-tuples
having $j=n+1/2$, can be expressed in the form, 
$$
f(r,\theta,\phi,\psi)=r^{n+1/2}e^{i\psi/2}y(\theta,\phi),
\quad\text{with}\quad
y(\theta,\phi)\in \bigoplus_{m=-j}^{+j}\langle j, m\rangle,
$$
whereas the components 
$f$ of the $2n +1$-tuples
having $j=n$, can be expressed in the form, 
$$
f(r,\theta,\phi)=r^{n}Y(\theta,\phi),
\quad\text{with}\quad
Y(\theta,\phi)\in \bigoplus_{m=-j}^{+j}\langle j, m\rangle.
$$

\medskip
\noindent
{\bf 12.5. Corollary.} The procedure that follows from the results above
to produce the harmonic wave functions for integer and
half-integer angular momentum states is this: Either,
\begin{enumerate}
\medskip
\item Go from $\langle 0,0\rangle$ to $\langle j,-j\rangle$ using
$\left(\text{\bf j}_1^{+-}\text{\bf j}_2^{+-}\right)^n$ if $j=n$,\\
(or using $\text{\bf j}_2^{+-}\left(\text{\bf j}_1^{+-}\text{\bf j}_2^{+-}\right)^n$
if $j=n+1/2$),\\
and then move from  $\langle j,-j\rangle$ to  $\langle j,+j\rangle$
by applying succesively $\text{\bf L}_+$; or,
 \medskip
 \item Go from $\langle 0,0\rangle$ to $\langle j,+j\rangle$ using
$\left(\text{\bf j}_2^{++}\text{\bf j}_1^{++}\right)^n$ if $j=n$,\\
(or using $\text{\bf j}_1^{++}\left(\text{\bf j}_2^{++}\text{\bf j}_1^{++}\right)^n$
 if $j=n+1/2$),\\ 
and then move from  $\langle j,+j\rangle$ to  $\langle j,-j\rangle$
by applying succesively $\text{\bf L}_-$.
\end{enumerate}
\medskip
\noindent
In the first case, each application of $\text{\bf L}_+$ changes 
the eigenvalue $m$ of $\text{\bf L}_z$
by adding $1$ in each step, whereas in the second case, 
each application of $\text{\bf L}_-$ changes $m$
by subtracting $1$ in each step. In particular,
\begin{enumerate}
\item[(3)]
Application of the operators $\text{\bf j}_*^{++}$ and $\text{\bf j}_*^{+-}$
to any of the above states
raise the eigenvalue $j$ by a $1/2$ step, whereas the operators 
$\text{\bf j}_*^{-+}$ and $\text{\bf j}_*^{--}$
lower $j$ by a $1/2$ step.
\end{enumerate}

\medskip
\noindent
\section*{13. Non–relativistic quantum equation for the 
Hydrogen atom including the electron spin}

\medskip
\noindent
As an example we now write a new
non–relativistic quantum equation (NNRQE) for the
Hydrogen atom that takes into account the electron spin.
The equation can be solved exactly in terms of the new
angular momentum operators depending on the Euler angle $\psi$
and the new harmonic wave functions introduced before.
The NNRQE is expressed in terms of four spatial variables plus time. 
The four spatial variables are the usual three–dimensional 
spherical coordinates
$r$, $\theta$, $\phi$, plus the third Euler angle $\psi$.

\medskip
\noindent
For comparison purposes, we use here
the standard units in terms of $\hbar$.
The NNRQE is then written in the usual fashion,
\begin{equation}\label{(65)}
\text{\bf H}\Psi(r,\theta,\phi,\psi)
=i\hbar\displaystyle{\frac{\partial \Psi(r,\theta,\phi,\psi)}{\partial t}},
\end{equation}
for the Hamiltonian operator $\text{\bf H}$ given by
\begin{equation}\label{(66)}
\text{\bf H} =
-\displaystyle{\frac{\hbar^2}{2\mu}}{\nabla_{4}}^2
+V(r,\theta,\phi,\psi).
\end{equation}
Here, $\mu$ is the reduced mass of the proton-electron system.
The only difference with the usual treatment is due to the presence of the new 
variable $\psi$, which implies that the four–dimensional Laplacian
${\nabla_{4}}^2(r,\theta,\phi,\psi,\partial_r,\partial_\theta,\partial_\phi,\partial_\psi)$
is related to the usual three–dimensional Laplacian
${\nabla_{3}}^2(r,\theta,\phi,\partial_r,\partial_\theta,\partial_\phi)$
by
\begin{equation}\label{(67)}
{\nabla_{4}}^2={\nabla_{3}}^2
+\left[\displaystyle{\frac{1}{r^2}}\left(
\csc^2\theta\displaystyle{\frac{\partial^2}{\partial\psi^2}}
-2\csc^2\theta\cot\theta
\displaystyle{\frac{\partial}{\partial\phi}}\displaystyle{\frac{\partial}{\partial\psi}}
\right)\right],
\end{equation}
where the correction (in square brackets) arises from the difference 
between the angular momentum operators in four and three dimensions
given in \eqref{(42)}.
Therefore, the Hamiltonian $\text{\bf H}$ written in full detail is
\begin{equation}\label{(68)}
\aligned
\text{\bf H} & =
-\displaystyle{\frac{\hbar^2}{2\mu}}\left(
\displaystyle{\frac{\partial}{\partial r}}\left(r^2\displaystyle{\frac{\partial}{\partial r}}\right)
+
\displaystyle{\frac{1}{\sin\theta}}\displaystyle{\frac{\partial}{\partial \theta}}
\left(\sin\theta \displaystyle{\frac{\partial}{\partial \theta}} \right) 
+
\displaystyle{\frac{1}{\sin^2\theta}}\displaystyle{\frac{\partial^2}{\partial\phi^2}}\right)
\\
&\quad
-\displaystyle{\frac{\hbar^2}{2\mu}}\left(
\displaystyle{\frac{1}{\sin^2\theta}}\displaystyle{\frac{\partial^2}{\partial\psi^2}}
-2\csc^2\theta\cot\theta
\displaystyle{\frac{\partial}{\partial\phi}}\displaystyle{\frac{\partial}{\partial\psi}}
\right)
+V(r,\theta,\phi,\psi).
\endaligned
\end{equation}

\medskip
\noindent
Consider the Coulomb potential for the Hydrogen atom,
\begin{equation}\label{(69)}
V(r,\theta,\phi,\psi)=-\displaystyle{\frac{q^2}{r}}.
\end{equation}
Observe that the four dimensional Hamiltonian 
$\text{\bf H}$ does not differ from the three dimensional one 
when acting on $\psi$-independent wavefunctions, so the usual solutions
for integer momentum $j$ (spinless electrons) are recovered.
There are, however, new exact solutions for the case of half–integer $j$
which solve exactly the NNRQE.
The wavefunction,
\begin{equation}\label{(70)}
\Psi_{j,j}(r,\theta,\phi,\psi)=Cr^j\,
e^{i\left(\frac{j\phi+\psi}{2}-\frac{Et}{\hbar}\right)-\frac{\mu q^2r}{(j+1){\hbar}^2}}\,
(\sin\theta)^{j-1/2}\cos\frac{\theta}{2}
\end{equation}
with energy,
\begin{equation}\label{(71)}
E_{j,j}=-\displaystyle{\frac{\mu q^4}{2{\hbar}^2(j+1)^2}},
\end{equation}
solves \eqref{(65)} for the Coulomb potential for any half-integer $j$ and $m=j$. 
It is straightforward to check the validity of these solutions 
by following the standard techniques of separation of variables;
one makes use of (2) in {\bf Cor. 12.5} and
a solution of the form $r^je^{a_jr}$ for the radial equation.
Finally, one may apply $\text{\bf L}_-$ to $\Psi_{j,j}$ repeatedly to get solutions 
$\Psi_{j,m}$ with the same energy $E_{j,m}=E_{j,j}$. 
This degeneracy is due to the $SO(4)$-symmetry
of the problem, as the NNRQE can be transformed into
an equation in the cartesian coordinates $u_1$, $u_2$, $u_3$ and $u_4$
of ${\mathbb R}^4$ introduced in \eqref{(9)} 
and its corresponding four-dimensional Laplace operator.
For the $SO(4)$-symmetry of the Hydrogen atom and the degeneration 
of its states see~\cite{Weinberg}.

\medskip
\noindent
\section*{14. Concluding Remarks}

\medskip
\noindent
We have given a new interpretation of the classical Hurwitz-Hopf map
$H:{\mathbb R}^4\to{\mathbb R}^3$, providing an assignment
${\mathbb C}^2\ni (z_1,z_2)\mapsto (r,\theta,\phi,\psi)$
that realizes the double covering map of groups, 
$G_0={\mathbb R}^+\times SU(2)\to {\mathbb R}^+\times SO(3)$,
keeping the real $3$-dimensional euclidean space interpretation
of both, the radius $r=\sqrt{{x_1}^2+{x_2}^2+{x_3}^2}$, and the Euler angles
$(\theta,\phi,\psi)$ that parametrize rotations in the group $SO(3)$.
The interpretation only requires that
$\left(\begin{smallmatrix} 
\bar{z}_1/\sqrt{r} & \,\,-\bar{z}_2/\sqrt{r} \\ z_2/\sqrt{r} & \,\,\,\,\, z_1/\sqrt{r}
\end{smallmatrix}\right)\in SU(2)$, whenever
$r=\vert z_1\vert^2+\vert z_2\vert^2$.
The Lie algebra generators,
$\{\text{\bf L}, \text{\bf L}_x, \text{\bf L}_y, \text{\bf L}_z\}$ 
of $\mathfrak{g}=\operatorname{Lie}(G_0)$,
written in terms of the complex variables $\{z_1,z_2,\bar{z}_1,\bar{z}_2\}$,
satisfy,
$$
{\text{\bf L}_x}^2
+{\text{\bf L}_y}^2
+{\text{\bf L}_z}^2
=
{\text{\bf L}}^2 + \text{\bf L}
+\displaystyle{\frac{r}{4}}\,\Delta,
$$
where $\Delta$ is the Laplace operator in the domain ${\mathbb R}^4\simeq{\mathbb C}^2$.
We use these facts to produce a 
Hilbert space $\mathcal H$ of harmonic wave functions 
---depending on the four real variables
$(r,\theta,\phi,\psi)$---
with which one can describe either
integer or half-integer angular momentum states.
The angular momentum operators
$\{\text{\bf L}_x, \text{\bf L}_y, \text{\bf L}_z\}$ coincide
with their usual expressions in the spherical coordinates $(r,\theta,\phi)$
when they act on states having integer angular momentum,
but they show and additional term with a derivative depending on the angle $\psi$
when acting on half-integer angular momentum states.
The total angular momentum operator $\text{\bf L}$ gets expressed
as $r\,\partial_r$.
Finally, following Schwinger (see \cite{6} and \cite{7}),
the $4$-dimensional Lie algebra of 
$G_0$ is coupled with two Heisenberg
Lie algebras of $2$-dimensional harmonic oscillators
generated by 
$\{z_1,z_2,\bar{z}_1,\bar{z}_2\}$ and their adjoints,
thus producing raising and lowering operators that
change the total angular momentum in half-units.
These two Heisenberg Lie algebras
intersect at the identity operator $\operatorname{Id}_{\mathcal H}$.
All the operators close either into a $9$-dimensional
Lie algebra or, leaving out only the operator $\operatorname{Id}_{\mathcal H}$,
into a $(4|8)$-dimensional Lie superalgebra.
Both algebraic structures can be faithfully represented in 
the Hilbert space $\mathcal H$ obtained
by the $L^2$-completion of the span of the subspace of complex polynomials
in the complex variables $(z_1,z_2,\bar{z}_1,\bar{z}_2)$
with the usual Hermitian product having the Hermite weight factor
$\operatorname{exp}\left(-1/2(\vert z_1\vert^2+\vert z_2\vert^2)\right)$.
Moreover, ${\mathcal H}$ can be decomposed in the form $\mathcal H=
{\mathcal H}_0\oplus {\mathcal H}_1$, 
where $\mathcal H_0$ contains all the
states having integer angular momentum (or {\it space of bosons\/})
and $\mathcal H_1$ contains all the
states having half-integer angular momentum (or {\it space of fermions\/}).
The closed subspaces ${\mathcal H}_0$ and ${\mathcal H}_1$ are obtained by 
producing first the highest weight vector
$\vert\,j\,,+j\,\rangle=(z_1\bar{z}_2)^j$ (if $j$ is integer, with $j\ge 1$)
or $\vert\,j\,,+j\,\rangle=z_1(z_1\bar{z}_2)^{\frac{2j-1}{2}}$ 
(if $j$ is half-integer, with $j\ge 1/2$),
and then move from it 
by successive application of the ladder operator $\text{\bf L}_-$
which preserves the eigenvalue $j$ but lowers the eigenvalue $m$ 
in one unit at each step, 
until reaching the lowest weight vector $\vert\,j\,,-j\,\rangle$,
characterized by $\text{\bf L}_-\vert\,j\,,-j\,\rangle=0$,
or by doing the analogue constructions
producing first the lowest weight vectors $\vert\,j\,,-j\,\rangle$.
As an example of this approach, a new non–relativistic quantum equation
for the hydrogen atom that includes the electron spin is given,
together with its eigensystem for half-integer angular momentum.

\bigskip

\medskip
\noindent
\section*{Acknowledgements}

\noindent
It is a pleasure to thank Octavio Casta\~nos--Garza and Luis F. Urrutia for fruitful conversations in the early stages of this work. Two of the authors acknowledge supports received by their research grants. 
EN-A acknowledges partial support by DGAPA-UNAM under project IN100323.
OA-SV acknowledges CONACYT Grant $\#$ A1-S-45886.


\medskip

\begin{thebibliography}{XXXX}
\addcontentsline{toc}{chapter}{References}

\bibitem{6} Schwinger J 1965 
{\it On Angular Momentum}, in
Quantum Theory of Angular Momentum (Biedenharn LC and Van Dam H, Eds.), 
Academic Press, New York, 229-279.

\bibitem{7} Schwinger J 1979 
{\it Multispinor Basis of Fermi-Bose Transformations\/,}
Ann. Phys. {\bf 119} 192-237.

\bibitem{Corwin75}
Corwin L, Ne'eman Y, and Sternberg S 1975 
Rev. Mod. Phys. {\bf 47} 573-603.

\bibitem{Freedman76}
Freedman DZ, Van Nieuwenhuizen P, and Ferrara S 1976 
Phys. Rev. D {\bf 13} 3214.

\bibitem{Deser76}
Deser S and Zumino B 1976 Phys. Lett. B {\bf 62} 335.

\bibitem{Schwinger66}
Schwinger J 1966 Phys. Rev. {\bf 152} 1219.

\bibitem{Varadarajan04}
Varadarajan VS 2004 {\it Supersymmetry for Mathematicians, An Introduction}, Courant Lecture Notes in Mathematics {\bf 11}, American Mathematical Society (New York, NY).\\
MR 2069561

\bibitem{Shun-Jen12}
Shun-Jen C and Weiqiang W 2012 {\it Dualities and Representations of Lie Superalgebras}, Graduate Studies in Mathematics {\bf 144}, American Mathematical Society (Providence, RI).\\
ISBN: 978-0-8218-9118-6

\bibitem{Jarvis23}
Jarvis P 2023 {\it Exploring the Basics of Lie Superalgebras and Their Applications in Mathematics and Physics}, J. Generalized Lie Theory App. {bf 17} (1) 372.\\
ISSN: 1736-4337

\bibitem{Cai22}
Cai ML, Wu YK, Mei QX, Zhao WD, Jiang Y, Yao L, He L, Zhou ZC, and Duan LM 2022 Nature Comm. {\bf 13} 3412.\\
doi: 10.1038/s41467-022-31058-0

\bibitem{Miri13}
Miri MA., Heinrich M, El-Ganainy R, and Christodoulides DN 2013 Phys. Rev. Lett. {\bf 110} 233902.\\
doi: 10.1103/PhysRevLett.110.233902

\bibitem{Hirokawa15}
Hirokawa M 2015 Quant. Stud. Math. Found. {\bf 2} 379–388.\\
doi: 10.1007/s40509-015-0041-y

\bibitem{Tomka15}
Tomka M, Pletyukhov M, and Gritsev V 2015 Sci. Rep. {\bf 5} 13097.\\
doi: 10.1038/srep13097

\bibitem{Efetov99}
Efetov K 1999 {\it Supersymmetry in Disorder and Chaos}, Cambridge University Press, (Cambridge, UK).\\
ISBN: 978-0521663823

\bibitem{Gharibyan21}
Gharibyan H, Hanada M, Honda M, and Liu J 2021 J. High. Energy Phys., {\bf 140} 2021.\\
doi: 10.1007/JHEP07(2021)140

\bibitem{3} Howe R 1985 
{\it Dual Pairs in Physics: Harmonic Oscillators, Photons, Electrons, and Singletons\/,}
Lectures in Applied Math. {\bf 21} 179-207.

\bibitem{4} Kashiwara M and Vergne M 1978
 {\it On the Segal-Shale-Weil Representations and Harmonic Polynomials\/,}
Inventiones Math. {\bf 44} 1-47.

\bibitem{2} Hage Hassan M and Kibler M 1991
{\it On Hurwitz Transformations\/.}
Proceedings of the Workshop {\lq\lq}Le probl\`eme de factorisation de Hurwitz: approche historique, solutions, applications en physique{\rq\rq}, Eds., A. Ronveaux and
D. Lambert (Facult\'es Universitaires N.D. de la Paix, Namur) 1-29.
http://hal.in2p3.fr/in2p3-00115773/

\bibitem{Peniche-SV} Peniche R and S\'anchez-Valenzuela OA 2005
{\it Lie supergroups supported over $GL_2$ and $U_2$ associated to the adjoint representation\/,}
Journal of Geometry and Physics. {\bf 56} 999-1028

\bibitem{Rothstein} Rothstein M 1993
{\it Equivariant splittings of supermanifolds\/,} 
Journal of Geometry and Physics. {\bf 12} 145-152

\bibitem{5} Salgado G and S\'anchez-Valenzuela OA 2011
{\it Lie superalgebras over $\mathfrak{gl}_2$\/,}
Communications in Algebra. {\bf  39} 2114-2136.

\bibitem{Weinberg}
Weinberg, SJ
{\it The $SO(4)$ Symmetry of the Hydrogen Atom\/}
(2011)
https://hep.uchicago.edu//~rosner/p342/projs/weinberg.pdf

\end{thebibliography}
\end{document}